\documentclass[12pt]{article}

\usepackage[margin=1in]{geometry}
\usepackage{xcolor}
\newcommand{\rev}[1]{\textcolor{black}{#1}}
\usepackage{amsmath,amssymb,amsfonts}
\usepackage{graphicx}
\usepackage{booktabs}
\usepackage{enumitem}
\usepackage{caption}
\usepackage{needspace}
\usepackage{subcaption}
\usepackage{pdflscape}
\usepackage[hidelinks]{hyperref}
\usepackage[numbers,sort&compress]{natbib}
\usepackage{float}
\usepackage[none]{hyphenat}
\usepackage[table]{xcolor}
\usepackage{multirow}
\title{Interfacial Accommodation as a Candidate Ductility Pathway in Intermetallic-Rich Alloys}

\author{Avik Mahata\\
  Department of Mechanical and Electrical Engineering\\
  Merrimack College, North Andover, MA 01845, United States\\
  \texttt{mahataa@merrimack.edu}}

\date{}

\begin{document}
\maketitle

%% ---------------------------------------------------------------
%%  ABSTRACT 
%% ---------------------------------------------------------------

\begin{abstract}
\rev{
Heterophase interfaces are increasingly recognized as active participants in plastic deformation, yet quantitative methods for comparing their accommodation capacity across different interface chemistries and crystallographies remain limited. Here, we present an atomistic framework for quantifying interface-mediated strain accommodation based on molecular dynamics simulations. Interface broadening, roughening, and migration are extracted directly from atomistic trajectories and combined into an Interface Accommodation Index,
$A_{\mathrm{int}}=\Delta W+\Delta R+|\Delta z|$,
with a normalized counterpart,
$D_{\mathrm{int}}=\Delta W/W_0+\Delta R/R_0$,
that accounts for differences in the initial interface structure. Within a disconnection-based interpretation, these observables provide complementary measures of interface structural accommodation, while a weighting sensitivity analysis demonstrates that the relative ranking of interfaces is robust to the specific form of the index. The framework is demonstrated for three experimentally motivated interfaces, Al/Al$_3$Ti, Al/Al$_9$M$_2$, and Al$_9$M$_2$/Al$_3$Ti (M = Fe, Co, Ni), identified in additively manufactured aluminum alloy, under tensile, compressive, and shear loading. Bulk simulations show that Shockley partial dislocations dominate plastic deformation, with alloy chemistry governing the transition toward mixed-character dislocation networks. Among the interfaces examined, tensile loading produces the greatest accommodation response, whereas shear produces comparatively limited structural evolution. The Al$_9$M$_2$/Al$_3$Ti interface exhibits both the highest yield resistance and the largest accommodation index, indicating that intermetallic--intermetallic interfaces can simultaneously sustain load and redistribute strain. The simulations quantify structural accommodation at interfaces rather than ductility or fracture directly; the proposed connection to macroscopic ductility therefore remains a hypothesis requiring experimental validation. More broadly, the proposed framework provides a transferable methodology for comparing interface accommodation across multiphase alloy systems and identifies interface chemistry and crystallography as important design variables for developing damage-tolerant structural materials.
}
\end{abstract}

%% ---------------------------------------------------------------
%%  1. INTRODUCTION 
%% ---------------------------------------------------------------
\section{Introduction}

The development of structural alloys that simultaneously exhibit high strength and substantial ductility remains a central challenge in materials science. Additive manufacturing (AM), selective laser melting (SLM), and other rapid-solidification routes provide unique opportunities for addressing this challenge by generating nonequilibrium microstructures that are inaccessible through conventional processing \cite{DebRoy2021,Martin2017,Rometsch2022}. The extreme thermal gradients and rapid solidification conditions associated with AM promote the formation of nanoscale intermetallic phases, hierarchical microstructures, and high densities of interfaces that strongly influence mechanical behavior \cite{George2019, Miracle2017, kanjilal2025microscale, onuike2018additive, wu2020enhanced}. In Al-rich multicomponent alloys, these microstructures frequently consist of interconnected networks of ordered intermetallic compounds embedded within a ductile matrix. For example, additively manufactured Al$_{92}$(TiFeCoNi)$_2$ develops nanoscale Al$_3$Ti and Al$_9$(Fe,Co,Ni)$_2$ colonies that exhibit compressive strengths exceeding 700 MPa while retaining measurable plastic deformability \cite{Shang2024NatComm,Shang2024AM}. Similar strength--ductility synergies have been reported in compositionally complex alloys, precipitation-strengthened high-entropy alloys, and dual-phase systems where coherent and semi-coherent interfaces govern the interaction between dislocations and ordered phases \cite{Li2022CoNiV,Ding2019,Sohn2019}. \rev{A broader body of work on heterostructured and compositionally complex materials reinforces the view that interfaces, rather than bulk phases alone, control the attainable strength--ductility balance. Multiscale heterostructure engineering in large thickness-ratio Ti/Al/Mg clad plates has been shown to improve comprehensive properties through controlled interfacial architecture during heterothermal rolling \cite{Wei2025JMA}, while additively manufactured cryogenic chemically complex alloys develop sponge bone-like reticular nanoscale superstructures whose internal interfaces govern low-temperature deformability \cite{Xie2025CompB}. At higher length scales, interfacial wetting behaviour and the formation of intermediate reaction layers have been shown to control interface stability in high-temperature metallurgical systems \cite{Xu2025MMTB}. Interfacial chemistry has likewise been exploited deliberately as a design variable: grain-boundary carbon segregation and its transition to carbide precipitation modulates the strength--ductility trade-off in refractory high-entropy alloys \cite{Luo2026IJP}, and solubilizing Ta within the Cr$_2$Nb Laves phase confers remarkable high-temperature softening resistance in Cu--Cr--Nb alloys \cite{Shu2026SM}. Taken together, these studies establish interface density, interface chemistry, and interfacial phase selection as primary levers for mechanical performance, and provide the wider context within which the present interface-accommodation analysis is framed.} Despite these advances, the origin of ductility in intermetallic-rich alloys remains poorly understood. Intermetallic phases are traditionally viewed as strengthening constituents that impede dislocation motion while simultaneously promoting stress localization and brittle failure. Consequently, understanding how interfaces respond during deformation has become increasingly important for explaining the mechanical behavior of these materials.

Most existing studies interpret deformation in intermetallic-rich alloys through conventional bulk mechanisms, including dislocation transmission, antiphase-boundary formation, precipitate shearing, and Orowan looping \cite{kikuchi1979theory,allen1979microscopic,hazzledine1974coplanar,Ardell1985,Nembach1997}. However, recent experimental observations suggest that interfaces themselves may actively participate in strain accommodation. High-resolution transmission electron microscopy of deformed Al$_{92}$(TiFeCoNi)$_2$ reveals dislocations within Al$_3$Ti, stacking faults within Al$_9$(Fe,Co,Ni)$_2$, curved intermetallic lamellae, localized lattice rotations, and microcracks that remain confined within individual intermetallic domains rather than propagating catastrophically across colonies \cite{Shang2024NatComm}. Similar observations have been reported in nanostructured intermetallic systems and complex multiphase alloys where interfaces act as defect sources, sinks, and transmission pathways that redistribute plastic strain and suppress catastrophic fracture \cite{Xu2025,Rupert2019, zhang2023nanostructured, martin2021enhanced, xiao2025extraordinary, giwa2016microstructure, yang2015microstructural}. These findings suggest that interfaces are not merely passive barriers to deformation but dynamic structural elements whose evolution may directly influence ductility.

A useful framework for describing such behavior is provided by disconnection theory. Disconnections are interfacial line defects that possess both Burgers-vector and step character, enabling interfaces to accommodate shear while simultaneously migrating through the crystal \cite{HirthPond1996,Cahn2006,thomas2019disconnection}. Over the past two decades, disconnection-mediated mechanisms have successfully explained grain-boundary migration, shear-coupled motion, grain rotation, roughening transitions, and defect absorption across a wide range of crystalline materials \cite{Han2018,Frolov2013,Rajabzadeh2013,Zhu2019,Wei2019,Thomas2017,HanThomasSrolovitz2018}. Although these concepts are now well established for grain boundaries, their application to heterophase interfaces in intermetallic-rich alloys remains largely unexplored. In particular, it is unclear whether interfaces between matrix and intermetallic phases, or between distinct intermetallic compounds, can provide accommodation pathways analogous to those observed in grain boundaries. The present study builds upon a series of atomistic investigations of Al-based alloy solidification, defect evolution, and mechanical deformation conducted by Mahata et. al \cite{Mahata2018Nucleation,Mahata2019JCG}. Previous molecular dynamics studies established fundamental relationships between rapid solidification, defect formation, and subsequent mechanical response in Al and Al-alloy systems, including homogeneous nucleation, solidification defect evolution, deformation of nano-polycrystalline Al, and AM-relevant Al alloy solidification pathways \cite{Mahata2018Nucleation,Mahata2019JCG,Mahata2019CMS,Mahata2024MTC, mahata2026development}. Together with the development of Al-transition-metal interatomic potentials capable of accurately describing both thermodynamic and mechanical behavior \cite{Mahata2022MEAM}, these efforts provide the atomistic foundation for the present investigation of deformation accommodation at \rev{experimentally motivated} heterophase interfaces.

\rev{The novelty of this work does not reside in the concepts of interface-mediated plasticity or disconnection-mediated boundary motion, both of which are well established for grain boundaries in single-phase metals \cite{HirthPond1996,Cahn2006,Han2018}. Three specific contributions are made instead. First, disconnection-motivated accommodation analysis is extended from homophase grain boundaries to chemically complex heterophase interfaces separating a ductile FCC matrix from two structurally dissimilar intermetallics, and to the intermetallic--intermetallic boundary between them, a class of interface for which comparable analyses are not available. Second, a compact and directly measurable operator set is defined, consisting of the increase in interface width, or interface broadening ($\Delta W$), the increase in the root-mean-square fluctuation of the interface plane, or interface roughening ($\Delta R$), and the net displacement of the interface centroid along its normal, or interface migration ($\Delta z$). These three operators can be extracted per frame from any atomistic trajectory without requiring enumeration of individual disconnection modes, providing a transferable and reproducible basis for comparing interfaces of arbitrary chemistry and symmetry. Third, using this operator set we show that accommodation capacity and interfacial strength are not inversely correlated in these systems: the strongest interface examined is also the most accommodating. This last result is not anticipated by the conventional picture in which intermetallic interfaces act purely as obstacles, and it is the principal new physical finding reported here.}

In this work, we investigate the deformation behavior of three representative heterophase interfaces, Al/Al$_3$Ti, Al/Al$_9$M$_2$, and Al$_9$M$_2$/Al$_3$Ti (M = Fe, Co, Ni), \rev{whose phase pairings are taken from the intermetallic networks reported experimentally in} additively manufactured Al-rich alloys. \rev{It should be noted that it is the phase combinations, together with the orientation relationship of the intermetallic--intermetallic boundary reported from electron diffraction \cite{Shang2024NatComm}, that are experimentally established; the terminating planes of the two matrix--intermetallic boundaries were constructed to reproduce the observed rosette morphology, as detailed in Section~\ref{sec:interface_construction}. The models are therefore experimentally motivated rather than direct atomistic replicas of imaged boundaries.} \rev{Specifically, this study combines large-scale directional-solidification molecular dynamics, bulk mechanical simulations, and atomistically resolved heterophase interface models to establish the origin of deformation accommodation in intermetallic-rich alloys. The three experimentally motivated interfaces are characterised in terms of their crystallography, coherency, and interfacial properties, and subsequently subjected to tensile, compressive, and shear loading. Interface broadening ($\Delta W$), roughening ($\Delta R$), and migration ($\Delta z$) are extracted directly from the atomistic trajectories and combined into the Interface Accommodation Index $A_{\mathrm{int}}$ and its normalized counterpart $D_{\mathrm{int}}$. The resulting interface rankings are evaluated through a weighting sensitivity analysis and interpreted in the context of experimentally observed interfacial deformation features reported for Al$_{92}$(TiFeCoNi)$_2$ \cite{Shang2024NatComm}.} The results establish a quantitative framework linking interface structural evolution to deformation accommodation and \rev{indicate} that intermetallic interfaces can act as active carriers of plasticity rather than passive obstacles to deformation.

%% ---------------------------------------------------------------
%% 2. INTERFACE ACCOMMODATION FRAMEWORK
%% ---------------------------------------------------------------

\section{Interface Accommodation Framework}

Plastic deformation in intermetallic-rich alloys is traditionally interpreted in terms of bulk dislocation mechanisms such as transmission, pile-up, and precipitate shearing. However, recent experimental observations indicate that interfaces themselves can actively participate in strain accommodation through migration, defect absorption, structural disordering, and local shear transfer \citep{Shang2024NatComm,Xu2025}. Understanding how interfaces respond to applied deformation is therefore essential for establishing structure--property relationships in alloys containing dense populations of intermetallic phases.

A useful conceptual framework for describing such behavior is provided by disconnection theory. Disconnections are interfacial defects that possess both step and dislocation character and can therefore couple interface migration with shear deformation \citep{HirthPond1996,Cahn2006,Han2018,Qiu2024}. Within this framework, the motion of an interface is governed by the collective evolution of interfacial defects carrying Burgers vectors and step heights. For a disconnection mode $m$, the driving force per unit length may be expressed as

\begin{equation}
f_m = \tau b_m - p h_m - \nabla_s \gamma,
\label{eq:driving_force}
\end{equation}

where $\tau$ is the resolved shear stress, $p$ is the normal pressure, $b_m$ and $h_m$ are the Burgers vector and step height associated with mode $m$, and $\nabla_s\gamma$ is the local interfacial energy gradient. The resulting interface migration and shear are commonly described through disconnection fluxes and their associated step and Burgers vector content \citep{Cahn2006,Han2018,Qiu2024}. Direct identification and enumeration of disconnection modes in large-scale molecular dynamics simulations of chemically complex heterophase interfaces remains challenging. \rev{In the heterophase interfaces considered here, this difficulty is compounded by the low symmetry and large primitive cells of the intermetallic phases, which prevent the construction of the dichromatic pattern required for conventional disconnection-mode enumeration. The chemical disorder on the transition-metal sublattice further complicates the analysis by making individual defect cores difficult to identify unambiguously.} Rather than explicitly tracking disconnection populations, the present work adopts an observable-based framework that focuses on measurable structural signatures of interface accommodation. Specifically, we quantify changes in interface width, interface roughness, and interface position during deformation. These quantities represent experimentally accessible and atomistically resolvable manifestations of accommodation processes and provide a practical means of comparing deformation responses across distinct interfaces and loading modes. \rev{Throughout this paper we therefore describe the measured structural response as disconnection-consistent: the observables are those that disconnection activity would be expected to produce, but they do not by themselves constitute proof that disconnections are the operative carriers.}

\subsection{Interface accommodation metrics}
\label{sec:accommodation_metrics}

The interface width, $W$, is determined from the spatial distribution of atomic environments across the interface using a hyperbolic-tangent fit to the interfacial profile. The interface roughness, $R$, is defined as the root-mean-square fluctuation of the local interface position relative to its mean location. Interface migration is quantified by the displacement of the interface centroid, $\Delta z$.

To compare accommodation behavior across different interfaces and loading conditions, we define an Interface Accommodation Index (IAI),

\begin{equation}
A_{\mathrm{int}} = \Delta W + \Delta R + |\Delta z|,
\label{eq:IAI_def}
\end{equation}

where

\begin{equation}
\Delta W = W_f - W_0,
\end{equation}

\begin{equation}
\Delta R = R_f - R_0,
\end{equation}

and $W_0$, $R_0$, $W_f$, and $R_f$ denote the initial and final interface width and roughness, respectively. Within a disconnection-based interpretation, interface broadening reflects the accumulation and propagation of interfacial defects, roughening captures the development of step-like topological features, and migration corresponds to the net transport of interfacial defect content. Larger values of $A_{\mathrm{int}}$ therefore indicate interfaces that accommodate deformation through coupled migration, broadening, and roughening, whereas smaller values correspond to interfaces that remain relatively sharp and immobile.

To account for differences in initial interface structure, a normalized accommodation parameter is also introduced:

\begin{equation}
D_{\mathrm{int}} = \frac{\Delta W}{W_0} + \frac{\Delta R}{R_0}.
\label{eq:Dint}
\end{equation}
The quantity $D_{\mathrm{int}}$ provides a dimensionless measure of interface evolution and allows direct comparison among interfaces with different initial widths and roughness values. Interfaces exhibiting large values of $D_{\mathrm{int}}$ are interpreted as accommodation-dominated, while interfaces with small values remain structurally stable and may promote strain localization. Although $A_{\mathrm{int}}$ and $D_{\mathrm{int}}$ do not directly measure disconnection densities or Burgers-vector fluxes, they quantify the observable consequences of interface accommodation. The framework therefore complements existing disconnection theories by characterizing the net structural response of heterophase interfaces rather than resolving the underlying defect populations. This observable-based approach enables direct comparison of matrix--intermetallic and intermetallic--intermetallic interfaces and provides a quantitative measure of accommodation capacity that can be extracted directly from atomistic simulations. \rev{Values of $D_{\mathrm{int}}$ are reported alongside $A_{\mathrm{int}}$ for every interface--loading combination in Table~\ref{tab:operators}, so that the two measures may be compared directly.}

\subsection{Accommodation operators and disconnection processes}
\label{sec:operator_disconnection}

Although the operators $\Delta W$, $\Delta R$, and $\Delta z$ are defined here as structural observables, each can be interpreted within the framework of disconnection theory. Disconnections are line defects possessing both Burgers-vector and step character, and their glide provides the fundamental mechanism for shear-coupled interface migration and grain-boundary motion \cite{HirthPond1996,Cahn2006,Han2018, Qiu2024}. The operators introduced here are therefore not intended to identify individual disconnections directly, but rather to quantify the structural consequences expected from their collective activity while remaining measurable from atomistic trajectories. Broadening ($\Delta W$) is interpreted as the widening of the interfacial transition region associated with repeated disconnection activity and the redistribution of lattice disregistry over multiple atomic planes. Roughening ($\Delta R$) reflects the accumulation and spatial distribution of local ledges or step heights generated as disconnections propagate along the interface. For a population of disconnections of mode $m$ with step height $h_m$ and areal density $n_m$, the roughness follows approximately $R^2 \sim \sum_m n_m h_m^2$, so that $\Delta R$ represents the second moment of the resulting step-height distribution. Migration ($\Delta z$) instead characterizes the net transport of the interface and is associated with an imbalance in the flux of disconnections having opposite step signs. In the disconnection framework, this corresponds to the first moment of the step population,
\[
\Delta z \approx \sum_m n_m h_m,
\]
while the associated Burgers-vector content,
\[
\sum_m n_m b_m,
\]
gives rise to the accompanying shear-coupled boundary motion~\cite{Cahn2006,Han2018,Qiu2024}. These correspondences motivate the choice of accommodation operators but do not establish a unique inverse mapping. A given combination of $(\Delta W,\Delta R,\Delta z)$ may arise from multiple underlying disconnection populations $\{n_m,b_m,h_m\}$, and interface broadening may also result from stress-assisted chemical intermixing, local amorphization, or other structural disordering processes that do not require explicit disconnection activity. Likewise, recent atomistic simulations have shown that disconnection pile-up and pinning can generate inclined or incoherent interface segments and ultimately trigger crack nucleation, illustrating that several distinct defect processes may contribute to the same structural observables \cite{Hussein2026}. Accordingly, $A_{\mathrm{int}}$ is introduced as a phenomenological descriptor of interface structural evolution rather than as a direct measure of disconnection density or Burgers-vector flux. Throughout this work, the measured accommodation mechanisms are therefore interpreted as being consistent with disconnection-mediated processes rather than as direct demonstrations of disconnection-mediated plasticity.

\rev{\subsection{Weighting and sensitivity of the accommodation index}}
\label{sec:sensitivity}

\rev{Equation~(\ref{eq:IAI_def}) combines $\Delta W$, $\Delta R$ and $|\Delta z|$ with equal weight. Two points justify this choice. First, the summation is dimensionally consistent: all three operators are lengths measured in \AA{} along the interface normal, so no implicit unit conversion or scaling is concealed within the definition. Second, in the absence of a first-principles argument for preferring one accommodation channel over another, equal weighting is the deliberately agnostic choice; introducing tuned coefficients would embed an assumption about relative importance that the present data cannot support.}

\rev{Because equal weighting is nonetheless a convention, we verified that the resulting ranking of interfaces is not an artefact of it. A generalised index}
\rev{
\begin{equation}
A_{\mathrm{int}}(\mathbf{w}) = w_W\,\Delta W + w_R\,\Delta R + w_z\,|\Delta z|,
\qquad w_W + w_R + w_z = 3,
\label{eq:IAI_weighted}
\end{equation}
}
\rev{The robustness of the Interface Accommodation Index was evaluated using a family of physically motivated weighting schemes. These included equal weighting $(1,1,1)$, broadening-dominant $(2,0.5,0.5)$, roughening-dominant $(0.5,2,0.5)$, migration-dominant $(0.5,0.5,2)$, and the three limiting cases in which only $\Delta W$, only $\Delta R$, or only $|\Delta z|$ was retained. The normalized index $D_{\mathrm{int}}$ defined by Eq.~(\ref{eq:Dint}), which removes the influence of the initial interface structure, was included as an eighth variant. For each index definition, the seven interface--loading combinations were ranked and compared with the equal-weight ranking to evaluate the sensitivity of the accommodation ordering to the chosen metric. The results and their physical interpretation are presented in Section~\ref{sec:pathways} and Table~\ref{tab:sensitivity}.}

%% ---------------------------------------------------------------
%%  3. COMPUTATIONAL METHODOLOGY  
%% ---------------------------------------------------------------
%% ---------------------------------------------------------------
%% 2. COMPUTATIONAL METHODOLOGY
%% ---------------------------------------------------------------

\smallskip
\subsection{Atomistic simulations}
\label{sec:atomistic}

\begin{figure}[H]
  \centering
  \includegraphics[width=1\linewidth]{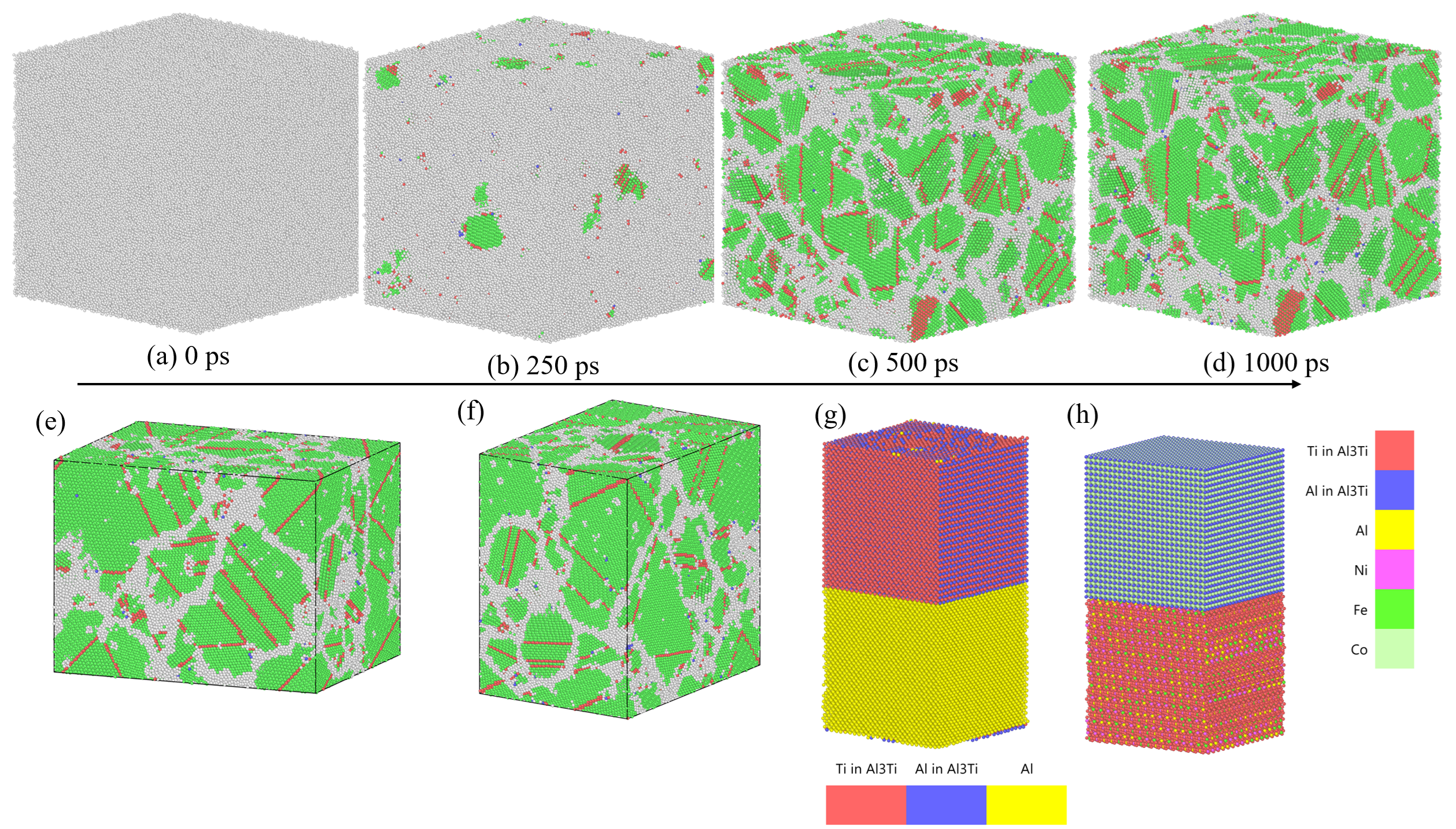}
 \caption{
Microstructure generation and interface models. (a--d) Solidification of the Al$_{92}$M$_2$ alloy (M = Ti, Fe, Co, Ni) from the liquid state to a polycrystalline intermetallic-rich microstructure. (e,f) Representative microstructures used for tensile and compressive deformation simulations. (g) Al/Al$_3$Ti interface model. (h) \rev{Al$_9$M$_2$/Al$_3$Ti interface model (M = Fe, Co, Ni).}
}
  \label{fig:microstructure}
\end{figure}

All molecular dynamics (MD) simulations were performed using LAMMPS \citep{Plimpton1995,Thompson2022LAMMPS}. Interatomic interactions were described using the modified embedded-atom method (MEAM) formalism originally developed by Baskes and subsequently extended by Lee and co-workers for multicomponent alloy systems \citep{Baskes1992MEAM,Lee2001MEAM}. For Al--Fe and Al--Ni interactions, the recently developed and validated MEAM potentials of Mahata \textit{et al.} were employed, which accurately reproduce thermodynamic, structural, and mechanical properties from room temperature to the melting point \citep{Mahata2022MEAM}. Remaining binary interaction terms were constructed following the standard 2NN--MEAM parameterizations of Baskes, Lee and co-workers \citep{Baskes1994HCP,Lee2003FCCMEAM}. \rev{A single five-element Al--Ti--Fe--Co--Ni library and parameter file was used for all simulations reported here. In the bicrystal models, the Al matrix and the Al sublattice of each intermetallic were assigned to Al, the Al$_3$Ti transition-metal sublattice to Ti, and the Al$_9$M$_2$ transition-metal sublattice to a random Fe/Co/Ni occupancy, allowing all phases and interfaces to be described consistently within a single potential library.} A timestep of 1~fs was used throughout. The overall deformation behaviour of Al-rich alloys was first investigated using large-scale directional-solidification simulations containing approximately $5\times10^{5}$ atoms within cubic cells of dimension $\sim$20~nm, covering compositions from pure Al to Al$_{86}$ in 2~at.\% increments with the balance distributed among Ti, Fe, Co, and Ni. The liquid alloy was equilibrated at 1200~K prior to solidification, and the resulting structures were subsequently equilibrated under an isothermal--isobaric (NPT) ensemble using a Nos\'{e}--Hoover thermostat and barostat \citep{Nose1984,Hoover1985,Martyna1994}. Figure~\ref{fig:microstructure}(a--d) illustrates the microstructural evolution from the homogeneous melt to a fully solidified polycrystalline aggregate containing intermetallic colonies, grain boundaries, and stacking-fault networks. Tensile and compressive loading were then applied along the $x$, $y$, and $z$ directions at an engineering strain rate of $10^{8}$~s$^{-1}$, consistent with standard practice in atomistic deformation studies. Loading along all three Cartesian directions minimises orientation bias arising from the arbitrary grain orientations produced during solidification. The solidification simulations identified three characteristic heterophase interfaces that capture the majority of interfacial interactions in these alloys: Al/Al$_3$Ti, Al/Al$_9$M$_2$, and Al$_9$M$_2$/Al$_3$Ti.\rev{ Here, M denotes Fe, Co, and Ni only. Ti partitions preferentially to Al$_3$Ti, consistent with the experimentally reported phase constitution of Al$_{92}$(TiFeCoNi)$_2$ \cite{Shang2024NatComm}, while the transition-metal sublattice of Al$_9$M$_2$ is occupied by Fe, Co, and Ni. Full structural and site-occupancy details are provided in Section~\ref{sec:phases}.} To isolate the role of each interface, atomistically resolved bicrystal models were constructed by generating orthogonal crystallographic slabs of the two constituent phases, replicating each slab to the target dimensions, and joining them within a periodic simulation cell while preserving crystallographic continuity across the interface. Representative models are shown in Fig.~\ref{fig:microstructure}(g,h). \rev{The interface planes, orientation relationships, lattice misfits, coherency state, relaxation protocol, and interfacial properties are summarized in Section~\ref{sec:interface_construction} and Table~\ref{tab:interfaces}.} Each bicrystal contained \rev{between $1.25$ and $1.32\times10^{5}$} atoms with the interface positioned near the cell centre and was subjected to uniaxial tension, uniaxial compression, and shear at a strain rate of $10^{9}$~s$^{-1}$. \rev{These strain rates are typical of molecular dynamics simulations and were applied consistently to every interface and loading condition. Consequently, the present work focuses on comparative trends among interfaces rather than direct quantitative comparison with laboratory strain rates.} Structural characterization was performed using OVITO \citep{Stukowski2010OVITO}. Common Neighbour Analysis (CNA) and Polyhedral Template Matching (PTM) were used to identify FCC, HCP, BCC, and disordered atomic environments, while dislocation evolution was quantified using the Dislocation Extraction Algorithm (DXA) \citep{Stukowski2012DXA}. Stress--strain responses, phase fractions, defect densities, stacking-fault populations, and atomic configurations were monitored continuously throughout each deformation simulation. \rev{The interface yield strengths reported in this work were determined using the conventional 0.2\% offset criterion to provide a consistent basis for comparison among the different interface configurations. Because the same definition was applied to every simulation, the reported values are intended for relative ranking of interface strength rather than direct correspondence with bulk engineering yield strengths.}

\rev{\subsection{Phase structures, site occupancy, and interatomic potential validation}}
\label{sec:phases}

\rev{Three crystalline phases are considered throughout this work: the FCC Al matrix ($Fm\bar{3}m$), the Al$_3$Ti intermetallic, and the medium-entropy Al$_9$M$_2$ intermetallic. Al$_3$Ti is represented by the metastable cubic L1$_2$ structure ($Pm\bar{3}m$, cP4), which has been reported in rapidly solidified and additively manufactured Al-rich alloys and is therefore adopted consistently in both the DFT calculations and the molecular dynamics simulations \cite{Shang2024NatComm}. The Al$_9$M$_2$ phase adopts the monoclinic Al$_9$Co$_2$ prototype ($P2_1/c$, mP22), the only crystallographic structure reported for the Al$_9$(Fe,Co,Ni)$_2$ intermetallic observed experimentally \cite{Shang2024NatComm}. The transition-metal sublattice is occupied by Fe, Co, and Ni in near-equiatomic proportion, while Ti partitions almost entirely to Al$_3$Ti, consistent with atom-probe measurements \cite{Shang2024NatComm}. Throughout this work, the notation Al$_9$M$_2$ therefore denotes the chemically disordered Fe--Co--Ni occupancy of the transition-metal sublattice. Interatomic interactions were described using the second-nearest-neighbour modified embedded-atom method (2NN--MEAM). The Al--Fe and Al--Ni interactions employed the recently developed potentials of Mahata \textit{et al.}, which were calibrated and validated from room temperature to the melting point \cite{Mahata2022MEAM}. The Al--Ti interaction was adopted from the 2NN--MEAM potential of Kim \textit{et al.} \cite{Kim2016TiAlMEAM}, while the Co--Al interaction employed the corresponding 2NN--MEAM potential developed by Dong \textit{et al.} \cite{Dong2012CoAlMEAM}. A single Al--Ti--Fe--Co--Ni potential library assembled from these validated binary interactions was used consistently throughout all bulk and interface simulations. The combined potential set was validated for every phase considered in this work. Table~\ref{tab:validation} compares lattice parameters, formation enthalpies, and representative elastic constants predicted by MEAM with DFT calculations and available experimental data. The lattice parameters of the three phases are reproduced within approximately 1--2\% of the reference values, while the calculated formation enthalpies remain negative for both intermetallic phases, consistent with first-principles predictions. Overall, the agreement demonstrates that the adopted potential library provides an accurate description of the equilibrium crystal structures and mechanical properties required for the atomistic simulations presented in this study. Beyond validating the interatomic potential, the first-principles calculations establish the thermodynamic basis of the phase constitution adopted in this work. Table~\ref{tab:dft} summarizes the total energies of the Al-rich compositions together with the four Al$_3$X prototype compounds. Among the ordered Al$_3$X structures, Al$_3$Ti possesses the lowest energy, supporting the preferential partitioning of Ti into the Al$_3$Ti phase while Fe, Co, and Ni remain in the Al$_9$M$_2$ intermetallic. Furthermore, decomposition of the chemically disordered Al-rich alloys into FCC Al and ordered intermetallic compounds is energetically favourable for every composition examined. These results independently confirm that the formation of Al$_3$Ti and Al$_9$M$_2$ colonies observed experimentally and reproduced during the directional-solidification simulations is thermodynamically expected rather than an artefact of the atomistic model. Finally, it should be noted that the present potential library is assembled from independently validated binary 2NN--MEAM interactions. No explicit reparameterization or validation of quaternary or quinary interaction terms was performed. Nevertheless, the excellent agreement with DFT, available experimental data, and the experimentally observed phase constitution indicates that the adopted potential is sufficiently transferable for the objectives of the present study. The development of dedicated multicomponent Al--Ti--Fe--Co--Ni potentials would nevertheless be valuable for future quantitative investigations of higher-order chemical interactions.}

\rev{
\begin{table}[H]
\centering
\caption{\rev{Validation of the 2NN--MEAM potential set for the three phases used in this study. MEAM values are compared with DFT and experimental reference data for Al \cite{simmons1971single}. Lattice parameters $a$ (and $b$, $c$, $\beta$ for the monoclinic phase) are in \AA{} and degrees; formation enthalpy $\Delta H_f$ in eV/atom; elastic constants and bulk modulus $B$ in GPa. Experimental Al$_9$M$_2$ parameters are those of the Al$_9$Co$_2$ prototype \cite{Shang2024NatComm}.}}
\label{tab:validation}
\small
\begin{tabular}{llccc}
\hline
Phase & Property & MEAM (this work) & DFT & Expt. \\
\hline
\multirow{5}{*}{Al (FCC, $Fm\bar{3}m$)}
 & $a$ (\AA)              & 4.05 & 4.05 & 4.05 \\
 & $C_{11}$ (GPa)         & 111.5 & 110 & 114.3 \\
 & $C_{12}$ (GPa)         & 58.8 & 61  & 62  \\
 & $C_{44}$ (GPa)         & 29.5 & 32  & 32  \\
\hline
\multirow{4}{*}{Al$_3$Ti (L1$_2$, $Pm\bar{3}m$)}
 & $a$ (\AA)              & 4.09 & 3.97 & -- \\
 & $\Delta H_f$ (eV/atom) & $-0.37$ & $-0.40$ & -- \\
 & $C_{11}$ (GPa)         & 102.76 & 129.50 & -- \\
 & $C_{44}$ (GPa)         & 38.33 & 26.20 & -- \\
  & $B$ (GPa)         & 87.18 & 83.80 & -- \\
\hline
\multirow{6}{*}{Al$_9$M$_2$ (mP22, $P2_1/c$)}
 & $a$ (\AA)              & 6.14 & -- & 6.22 \\
 & $b$ (\AA)              & 6.39 & -- & 6.29 \\
 & $c$ (\AA)              & 8.63 & -- & 8.56 \\
 & $\beta$ (deg)          & 94.6 & -- & 94.8 \\
 & $\Delta H_f$ (eV/atom) & $-0.29$ & $-0.33$ & -- \\
 & $C_{11}$ (GPa)         & 192.39 & 243.78 & -- \\
 & $C_{12}$ (GPa)         & 109.11 & 71.10 & -- \\
 & $C_{44}$ (GPa)         & 40.79 & 42.61 & -- \\
 & $B$ (GPa)              & 131.86 & 126.50 & -- \\
\hline
\end{tabular}
\end{table}
}

\rev{
\begin{table}[H]
\centering
\caption{\rev{First-principles total energies for the Al-rich compositions and Al$_3$X prototype compounds (spin-polarised DFT). $\Delta E_{\mathrm{decomp}}$ is the energy change on decomposing a chemically disordered Al-rich cell into FCC Al plus the ordered L1$_2$ Al$_3$X compounds; negative values indicate that ordering into intermetallic compounds is favourable. The Al$_3$X compounds are evaluated in the L1$_2$ structure used in the MD models.}}
\label{tab:dft}
\small
\begin{tabular}{llccc}
\hline
Structure & Prototype & Al (at.\%) & $E$ (eV/atom) & $\Delta E_{\mathrm{decomp}}$ (meV/atom) \\
\hline
Al                     & FCC   & 100   & $-3.741$ & -- \\
Al$_3$Ti               & L1$_2$ & 75   & $-5.118$ & -- \\
Al$_3$Fe               & L1$_2$ & 75   & $-5.004$ & -- \\
Al$_3$Co               & L1$_2$ & 75   & $-4.805$ & -- \\
Al$_3$Ni               & L1$_2$ & 75   & $-4.414$ & -- \\
Al$_{24}$Ti$_2$Fe$_2$Co$_2$Ni$_2$ & disordered & 75 & $-4.793$ & $-42.0$ \\
Al$_{80}$Ti$_5$Fe$_5$Co$_5$Ni$_5$ & disordered & 80 & $-4.585$ & $-30.9$ \\
Al$_{88}$Ti$_3$Fe$_3$Co$_3$Ni$_3$ & disordered & 88 & $-4.259$ & $-7.1$  \\
Al$_{92}$Ti$_2$Fe$_2$Co$_2$Ni$_2$ & disordered & 92 & $-4.084$ & $-7.3$ \\
\hline
\end{tabular}
\end{table}
}

\rev{\subsection{Heterophase interface construction and characterization}}
\label{sec:interface_construction}

\rev{Atomistically resolved bicrystal models were constructed for the three experimentally observed heterophase interfaces, Al/Al$_3$Ti, Al/Al$_9$M$_2$, and Al$_9$M$_2$/Al$_3$Ti. The objective was to reproduce the crystallographic interfaces reported in the additively manufactured Al$_{92}$(TiFeCoNi)$_2$ alloy rather than to perform an exhaustive survey of possible interface configurations. For the Al$_9$M$_2$/Al$_3$Ti interface, the orientation relationship was adopted directly from the transmission electron diffraction analysis of Shang \textit{et al.} \cite{Shang2024NatComm}, \[
[1\bar{3}0]_{\mathrm{Al_3Ti}}\parallel[100]_{\mathrm{Al_9M_2}},\qquad
(002)_{\mathrm{Al_3Ti}}\parallel(001)_{\mathrm{Al_9M_2}},\qquad
(310)_{\mathrm{Al_3Ti}}\parallel(010)_{\mathrm{Al_9M_2}},
\] for which the measured in-plane lattice misfits are 2.5\% and 2.1\%. For the Al/Al$_3$Ti and Al/Al$_9$M$_2$ interfaces, low-index terminating planes of the intermetallics were aligned with the corresponding low-index planes of the FCC Al matrix to reproduce the matrix--intermetallic interfaces observed within the intermetallic rosettes.} \rev{The in-plane lattice misfit was evaluated independently along the two interface directions using
\[
\delta_i=\frac{2(d_i^A-d_i^B)}{d_i^A+d_i^B},
\]
where $d_i^A$ and $d_i^B$ are the repeat distances of the two adjoining phases along the $i$th in-plane direction. Equilibrium lattice parameters predicted by the present potential were used to establish the commensurate supercells. A cube-on-cube $1{:}1$ registry was employed for the Al/Al$_3$Ti interface, whereas an approximate $3{:}2$ commensuration was adopted for the Al/Al$_9$M$_2$ interface. The remaining mismatch was accommodated by applying a small homogeneous strain to the Al-rich slab, thereby minimizing artificial coherency strain within the stiffer intermetallic phase. The resulting crystallographic orientation relationships, lattice misfits, coherency states, and geometric characteristics of all three interfaces are summarized in Table~\ref{tab:interfaces}. Each bicrystal was subsequently relaxed using a three-stage procedure. First, the structure was energy minimized using the conjugate-gradient algorithm with force and energy tolerances of $10^{-8}$. Second, the relative rigid-body translation of the two phases was optimized by systematically translating one slab over the in-plane periodic cell and minimizing the energy at each translation state to identify the lowest-energy interface registry. Finally, the optimized bicrystal was annealed under NPT conditions at 300~K for 100~ps and subjected to a second energy minimization to remove residual stresses and permit local interface reconstruction. All bicrystals satisfied zero-stress conditions prior to mechanical loading. The interfacial energy was then evaluated as}

\rev{
\begin{equation}
\gamma_{\mathrm{int}} = \frac{E_{\mathrm{tot}} - N_A e_A - N_B e_B}{2A},
\label{eq:gamma_int}
\end{equation}
}
\rev{where $E_{\mathrm{tot}}$ is the total energy of the relaxed bicrystal, $N_A$ and $N_B$ are the numbers of atoms belonging to each phase, $e_A$ and $e_B$ are the corresponding bulk cohesive energies per atom evaluated with the same potential, $A$ is the interface area, and the factor of two accounts for the two interfaces present in the periodic cell. The resulting orientation relationships, misfits, coherency states, and interfacial energies are collected in Table~\ref{tab:interfaces}.}

\rev{
\begin{table}[H]
\centering
\caption{\rev{Construction and characterization of the three heterophase interface models. Cell dimensions, atom counts, and interface area are obtained from the equilibrated bicrystal structures. The Al$_9$M$_2$/Al$_3$Ti orientation relationship and lattice misfit are taken from Shang \textit{et al.} \cite{Shang2024NatComm}. Reported interface areas correspond to a single interface in the periodic cell. Initial interface width $W_0$ and roughness $R_0$ are measured from the equilibrated pre-loading configurations.}}
\label{tab:interfaces}
\scriptsize
\setlength{\tabcolsep}{3pt}
\renewcommand{\arraystretch}{1.05}
\resizebox{\textwidth}{!}{%
\begin{tabular}{lccc}
\hline
Quantity & Al/Al$_3$Ti & Al/Al$_9$M$_2$ & Al$_9$M$_2$/Al$_3$Ti \\
\hline
Interface plane (phase A)      & $(001)_{\mathrm{Al}}$ & $(001)_{\mathrm{Al}}$ & $(002)_{\mathrm{Al_3Ti}}$ \\
Interface plane (phase B)      & $(001)_{\mathrm{Al_3Ti}}$ & $(001)_{\mathrm{Al_9M_2}}$ & $(001)_{\mathrm{Al_9M_2}}$ \\
Orientation relationship       & $[100]_{\mathrm{Al}}\!\parallel\![100]_{\mathrm{Al_3Ti}}$ & $[100]_{\mathrm{Al}}\!\parallel\![100]_{\mathrm{Al_9M_2}}$ & $[1\bar{3}0]_{\mathrm{Al_3Ti}}\!\parallel\![100]_{\mathrm{Al_9M_2}}$ \\
In-plane misfit $\delta_1$ (\%) & 1.0 & 3.6 & 2.5 \cite{Shang2024NatComm} \\
In-plane misfit $\delta_2$ (\%) & 1.0 & 5.0 & 2.1 \cite{Shang2024NatComm} \\
Coherency state                & semi-coherent & semi-coherent & semi-coherent \\
Cell dimensions (\AA{} $\times$ \AA{} $\times$ \AA{}) & $100.00 \times 100.00 \times 200.45$ & $100.00 \times 100.00 \times 201.30$ & $100.00 \times 100.00 \times 201.30$ \\
Interface area (\AA$^2$)       & $1.0 \times 10^{4}$ & $1.0 \times 10^{4}$ & $1.0 \times 10^{4}$ \\
Number of atoms                & $128{,}826$ & $131{,}179$ & $135{,}005$ \\
Initial width $W_0$ (\AA)      & $4.63$ & $5.29$ & $96.80$ \\
Initial roughness $R_0$ (\AA)  & $0.38$ & $0.47$ & $1.18$ \\
\hline
\end{tabular}%
}
\end{table}
}

\subsection{Interface accommodation and mechanics framework}

The present work is motivated by disconnection-based descriptions of interface motion, in which interfacial defects possessing both Burgers vector and step character couple interface migration with shear deformation \citep{HirthPond1996,Cahn2006,Han2018,Qiu2024}. While direct determination of disconnection densities, Burgers-vector fluxes, and individual disconnection modes remains challenging in large-scale molecular dynamics simulations of heterophase interfaces, the structural consequences of these processes can be quantified directly from atomistic trajectories. Accordingly, the present framework focuses on measurable descriptors of interface accommodation that provide a bridge between atomistic deformation mechanisms and continuum descriptions of interface-mediated plasticity. Interface evolution was characterized using three primary quantities: interface migration, interface broadening, and interface roughening. For each simulation frame, atom-type and structural-order distributions were evaluated along the interface normal direction. The interface centroid position, $z_0$, was determined from the midpoint of the compositional transition region, while the interface width, $W$, was obtained from a hyperbolic-tangent fit to the corresponding concentration profile. Interface roughness, $R$, was calculated as the root-mean-square fluctuation of the local interface position across the interface plane. These quantities provide direct measures of how an interface evolves during deformation and whether strain accommodation occurs through localized defect activity or distributed interfacial restructuring. The changes in interface width, roughness, and position during deformation are defined as

\begin{equation}
\Delta W = W_f - W_0,
\label{eq:deltaW}
\end{equation}

\begin{equation}
\Delta R = R_f - R_0,
\label{eq:deltaR}
\end{equation}

\begin{equation}
\Delta z = z_f - z_0,
\label{eq:deltaz}
\end{equation}

where the subscripts $0$ and $f$ denote the initial and final states, respectively. Here, $\Delta W$ quantifies interface broadening, $\Delta R$ characterizes the development of interface roughness, and $\Delta z$ represents the net migration of the interface during deformation.

To facilitate direct comparison among interfaces and loading modes, these quantities were combined into an Interface Accommodation Index (IAI),

\begin{equation}
A_{\mathrm{int}}
=
\Delta W
+
\Delta R
+
\lvert \Delta z \rvert,
\label{eq:IAI}
\end{equation}

which represents the overall magnitude of structural accommodation occurring at an interface. Larger values of $A_{\mathrm{int}}$ correspond to interfaces that accommodate deformation through migration, roughening, and broadening, whereas smaller values indicate interfaces that remain relatively immobile and structurally sharp. Although $A_{\mathrm{int}}$ is not a constitutive parameter, it provides a convenient quantitative descriptor for ranking interfaces according to their accommodation capacity. The analysis was performed in two stages. First, per-frame interface trajectories were extracted from the molecular dynamics simulations, yielding time-resolved measurements of interface position, width, roughness, local structural fractions, dislocation content, and slip activity. Second, these atomistic observables were post-processed to construct an interface operator database containing migration distances, broadening metrics, roughness evolution, accommodation indices, stress--strain descriptors, and deformation classifications. The resulting database enables systematic comparison of deformation accommodation pathways across matrix--intermetallic and intermetallic--intermetallic interfaces and provides the quantitative basis for the mechanistic analysis presented in the following sections. Within this framework, interface migration, roughening, and broadening are interpreted as observable signatures of interfacial accommodation that may arise from defect absorption, defect emission, and disconnection-mediated restructuring. The methodology therefore provides a physically motivated route for connecting atomistic deformation behavior to interface mechanics without requiring explicit enumeration of individual disconnection modes.

%% ---------------------------------------------------------------
%%  4. RESULTS AND DISCUSSION  
%% ---------------------------------------------------------------
\section{Results and Discussion}

\subsection{Composition-dependent deformation mechanisms}

Figure 2 summarizes the mechanical response and defect evolution of Al-rich alloys containing varying concentrations of transition-metal solutes. Significant differences in both strength and dislocation activity are observed as the alloy composition changes. The stress--strain curves in Figures~\ref{fig:ss}(a,b) show that all compositions undergo an initial elastic regime followed by yielding and plastic flow. The measured yield stresses exhibit a clear compositional dependence, with transition-metal-rich alloys generally displaying higher strength under both tensile and compressive loading (Figure~\ref{fig:ss}(c)). The difference between tensile and compressive yield stresses indicates an asymmetry in defect activation and evolution under the two loading modes. Direct examination of the dislocation structures reveals that plastic deformation is dominated by the nucleation, multiplication, and interaction of extended dislocation networks. Representative DXA snapshots (Figures~\ref{fig:ss}(d--f)) show a progressive increase in network complexity with increasing strain. Initially, the defect structure consists primarily of isolated dislocation segments. With continued deformation, these segments multiply, interact, and form interconnected networks that span large regions of the simulation cell. Analysis of the Burgers vector populations indicates that Shockley partial dislocations constitute the dominant deformation carriers across all compositions. However, the relative contribution of different dislocation families changes systematically with alloy chemistry. Alloys containing larger fractions of transition-metal elements exhibit a higher proportion of Shockley partials, consistent with deformation dominated by extended dislocation activity and stacking-fault formation. In contrast, increasing Al content promotes a gradual increase in perfect dislocations and complex defect structures, including sessile junctions and mixed Burgers-vector configurations. The growing fraction of these defects suggests enhanced dislocation interactions and network complexity in the Al-rich compositions. These observations demonstrate that alloy composition governs not only the mechanical response but also the nature of the defect structures responsible for plastic deformation. The transition from Shockley-partial-dominated plasticity toward increasingly complex mixed dislocation networks provides an atomistic explanation for the composition dependence of mechanical behavior in these Al-rich alloys. To quantify the evolution of these defect populations, strain-dependent dislocation density and Burgers-vector statistics are analyzed in the following section.

\begin{figure}[H]
  \centering
  \includegraphics[width=0.95\linewidth]{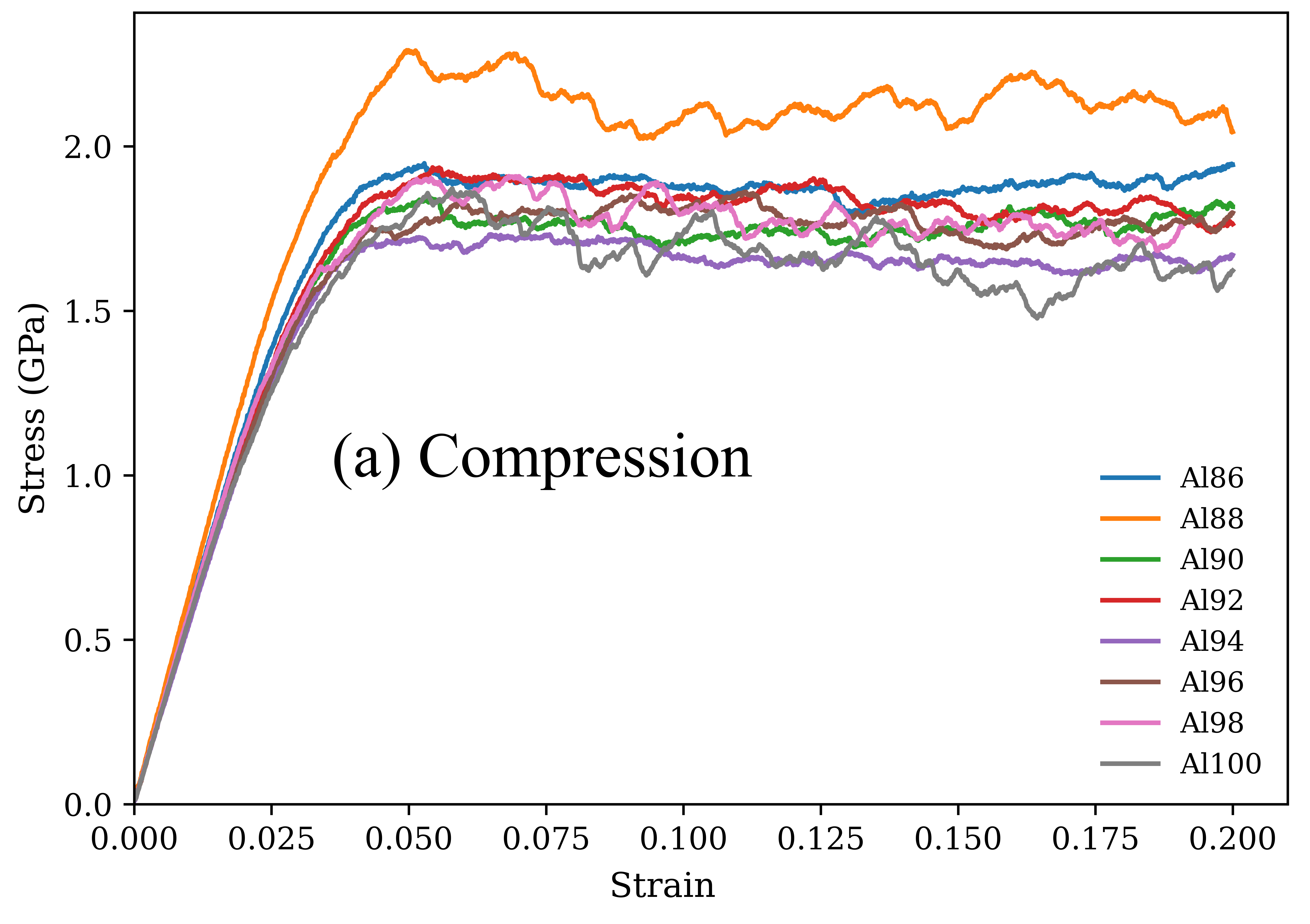}
  \includegraphics[width=0.95\linewidth]{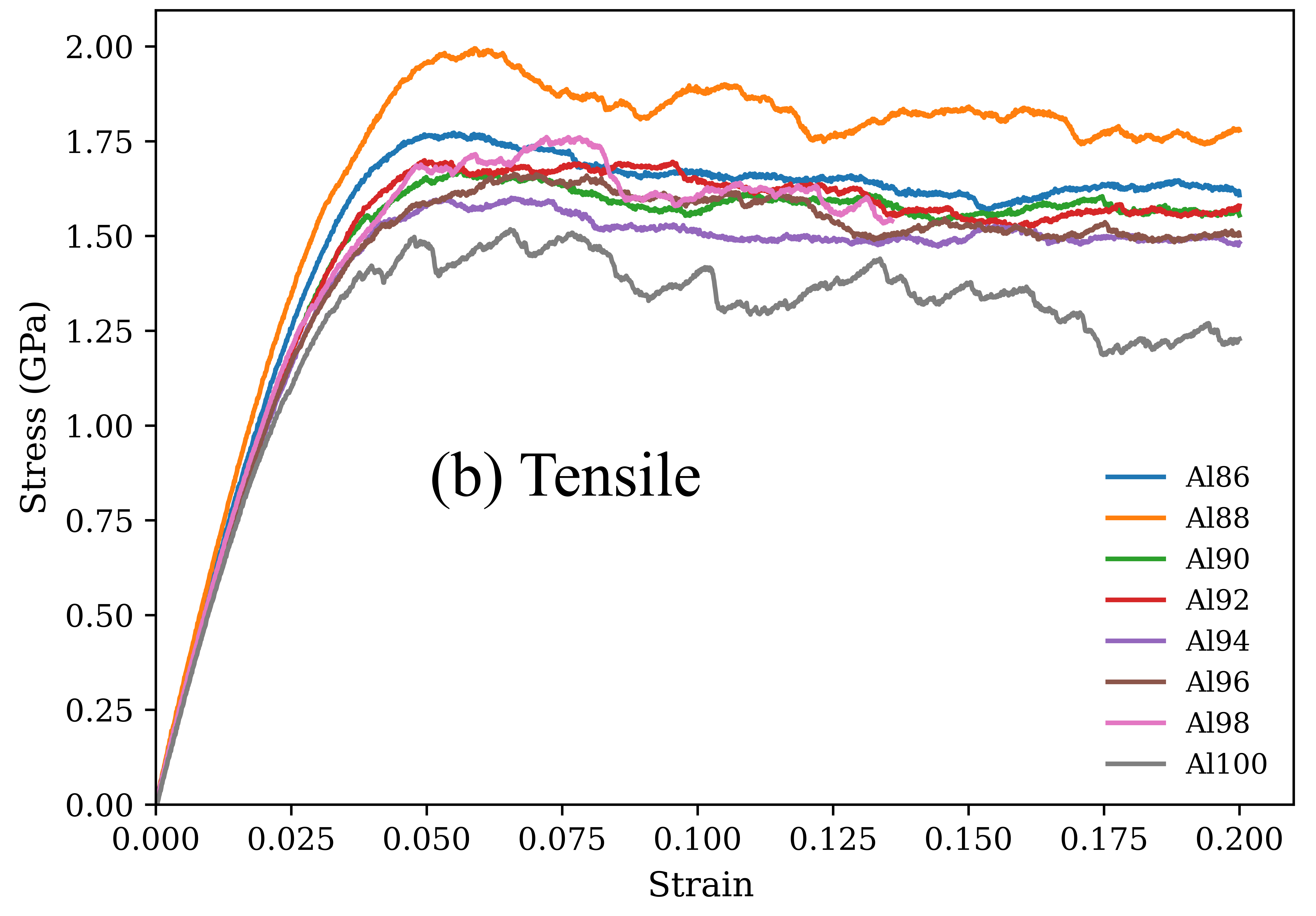}
\end{figure}

\begin{figure}[H]
  \centering
  \includegraphics[width=0.75\linewidth]{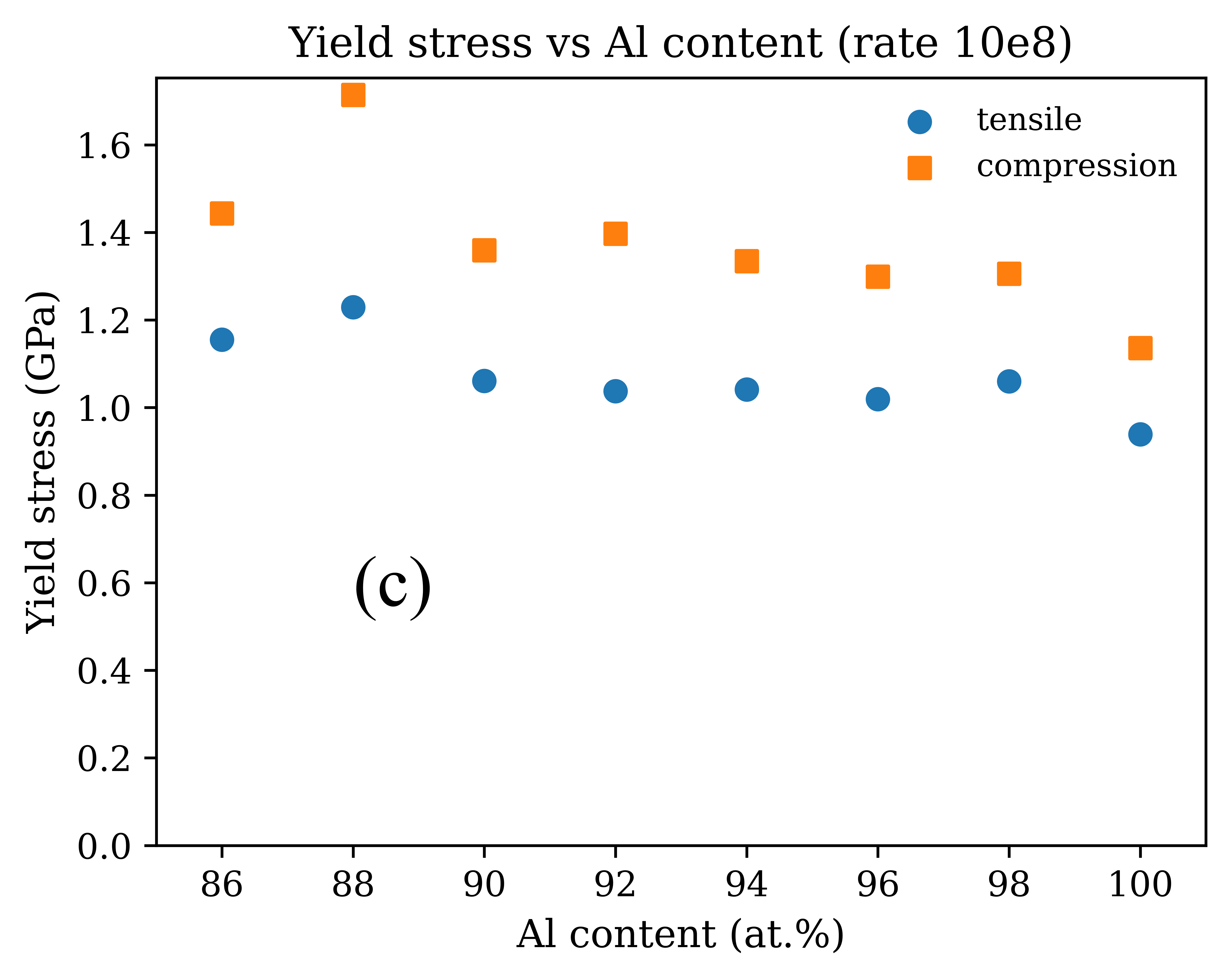}
  \includegraphics[width=0.75\linewidth]{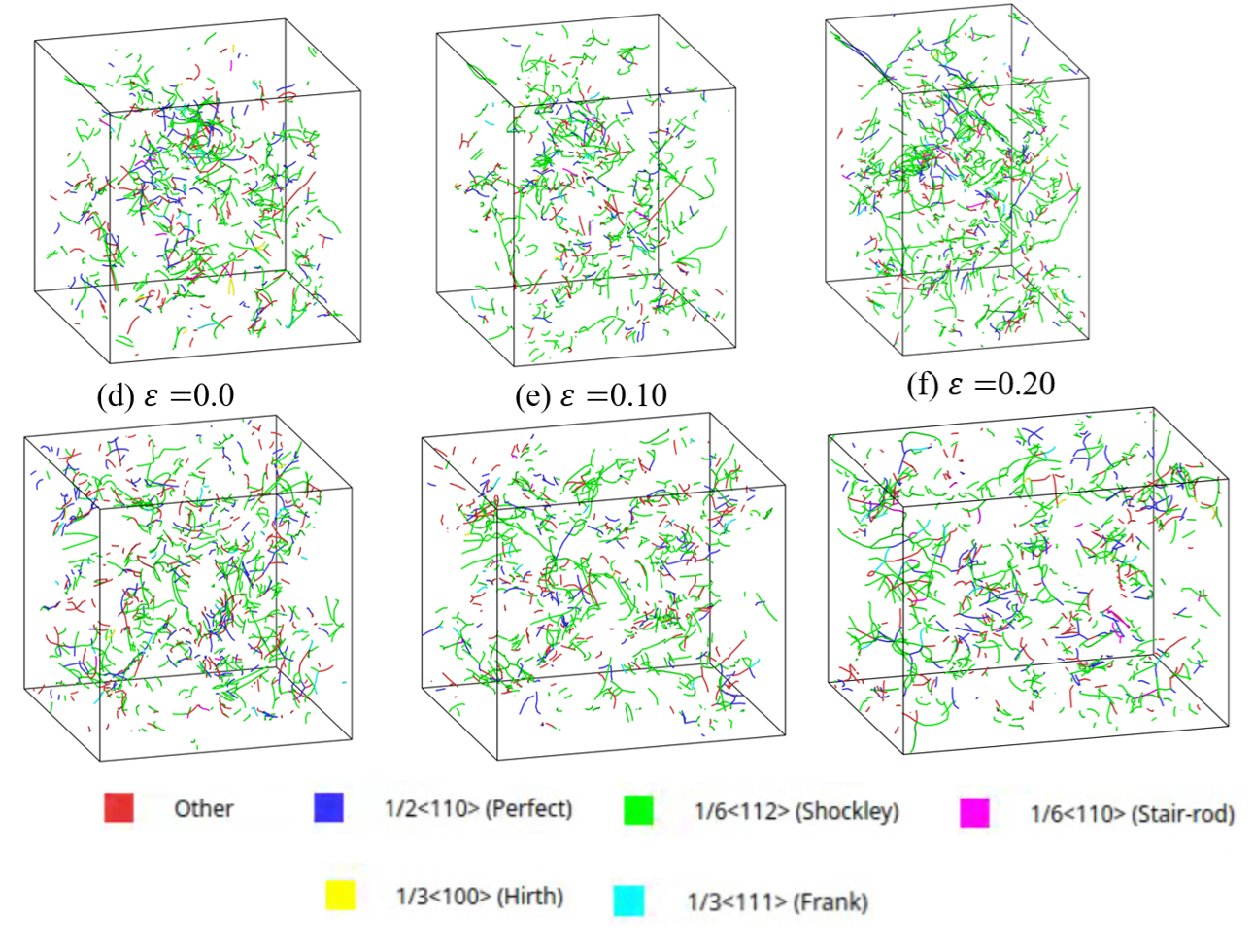}
  \caption{Mechanical response and dislocation evolution in Al-rich alloys under uniaxial loading at a strain rate of $10^{8}$ s$^{-1}$. (a) Compressive stress--strain curves and (b) tensile stress--strain curves for alloys containing 86--100 at.\% Al. (c) Yield stress as a function of Al content under compression and tension. (d--f) Representative DXA dislocation structures at engineering strains $\varepsilon = 0.0$, $0.10$, and $0.20$ during compressive deformation (top row) and tensile deformation (bottom row). Colors denote different Burgers vector families identified by dislocation extraction analysis.}
  \label{fig:ss}
\end{figure}

\subsection{Stress--strain response and defect evolution}

\begin{figure}[H] \centering \includegraphics[width=.65\linewidth]{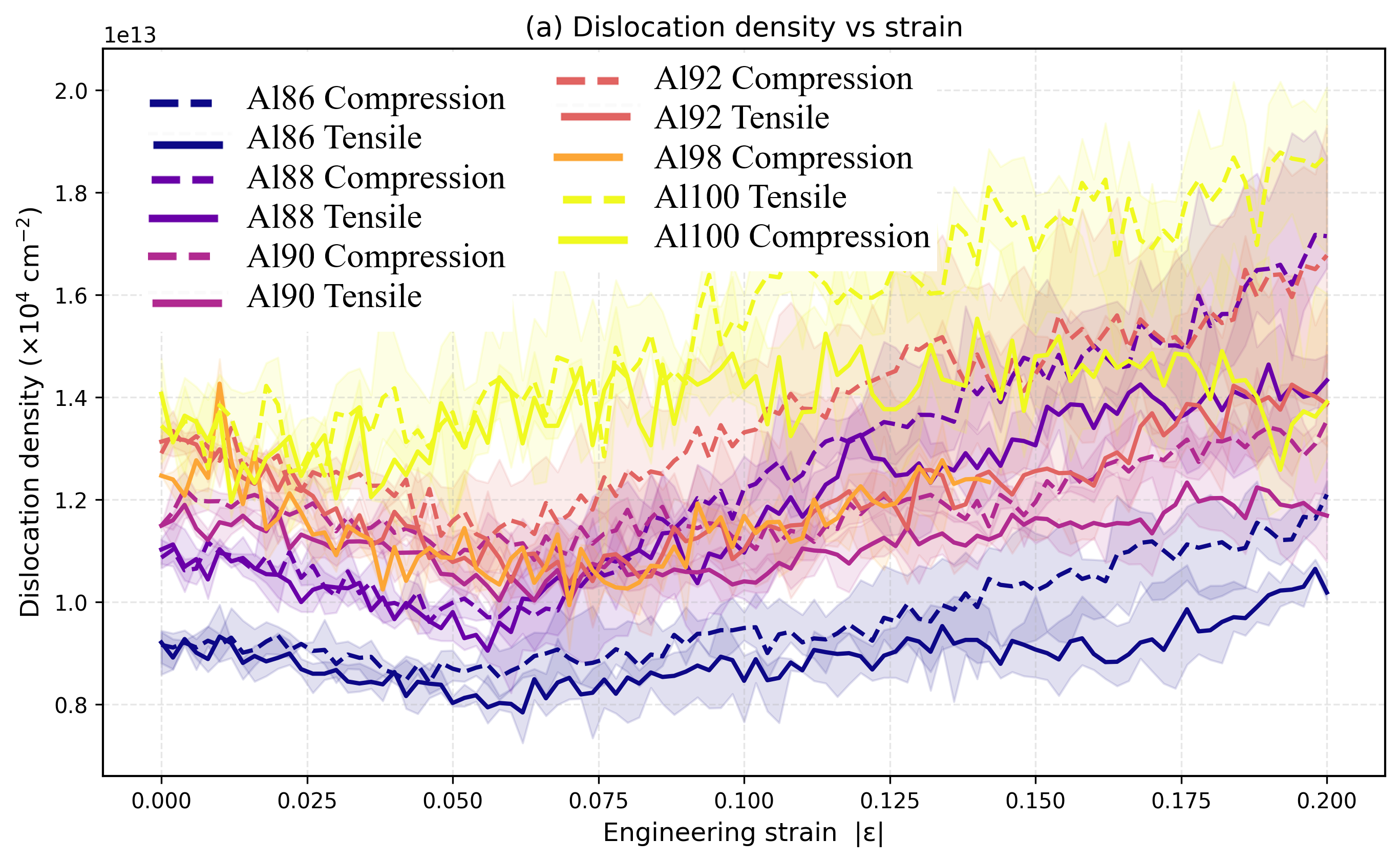} \includegraphics[width=.65\linewidth]{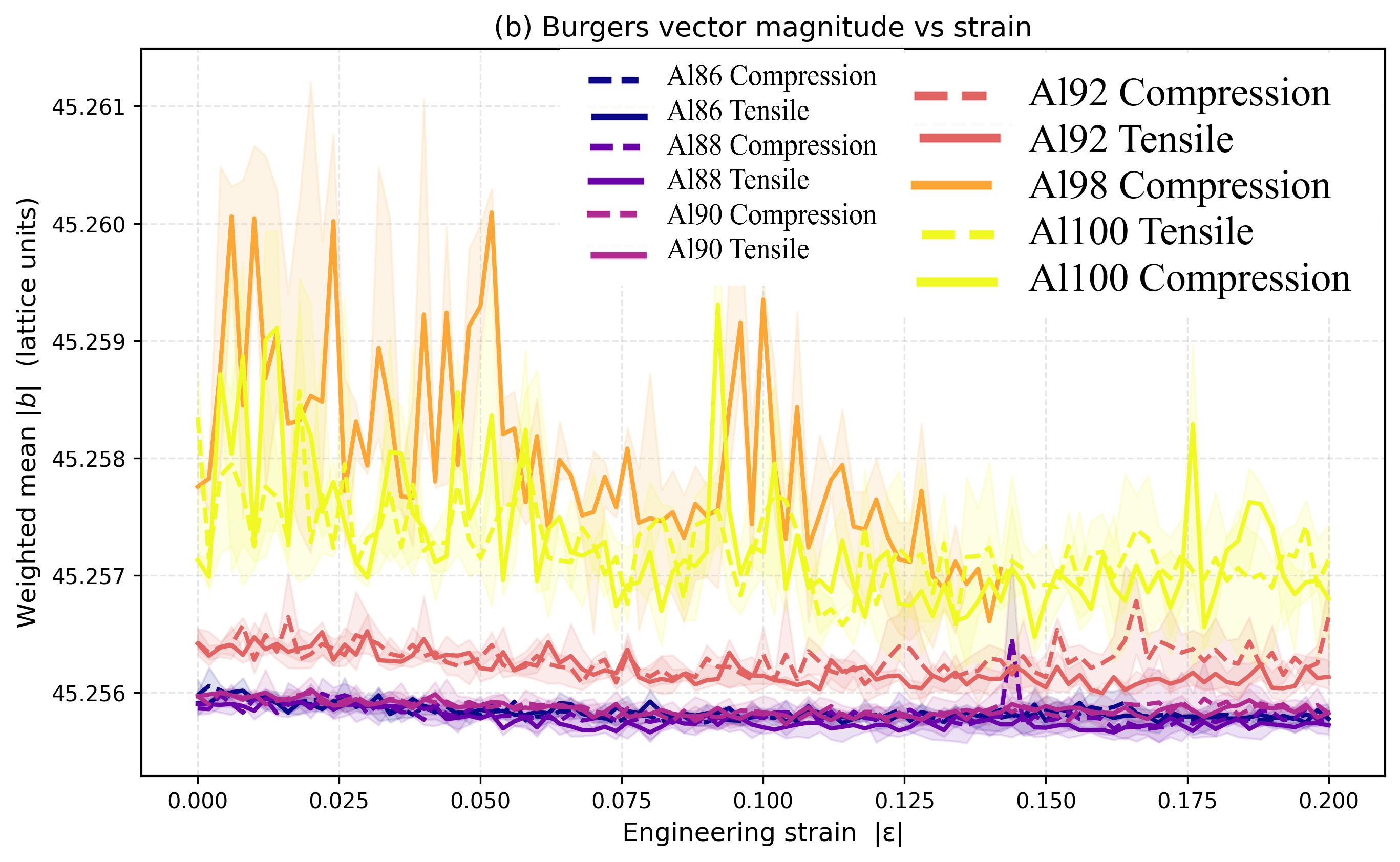} \includegraphics[width=.65\linewidth]{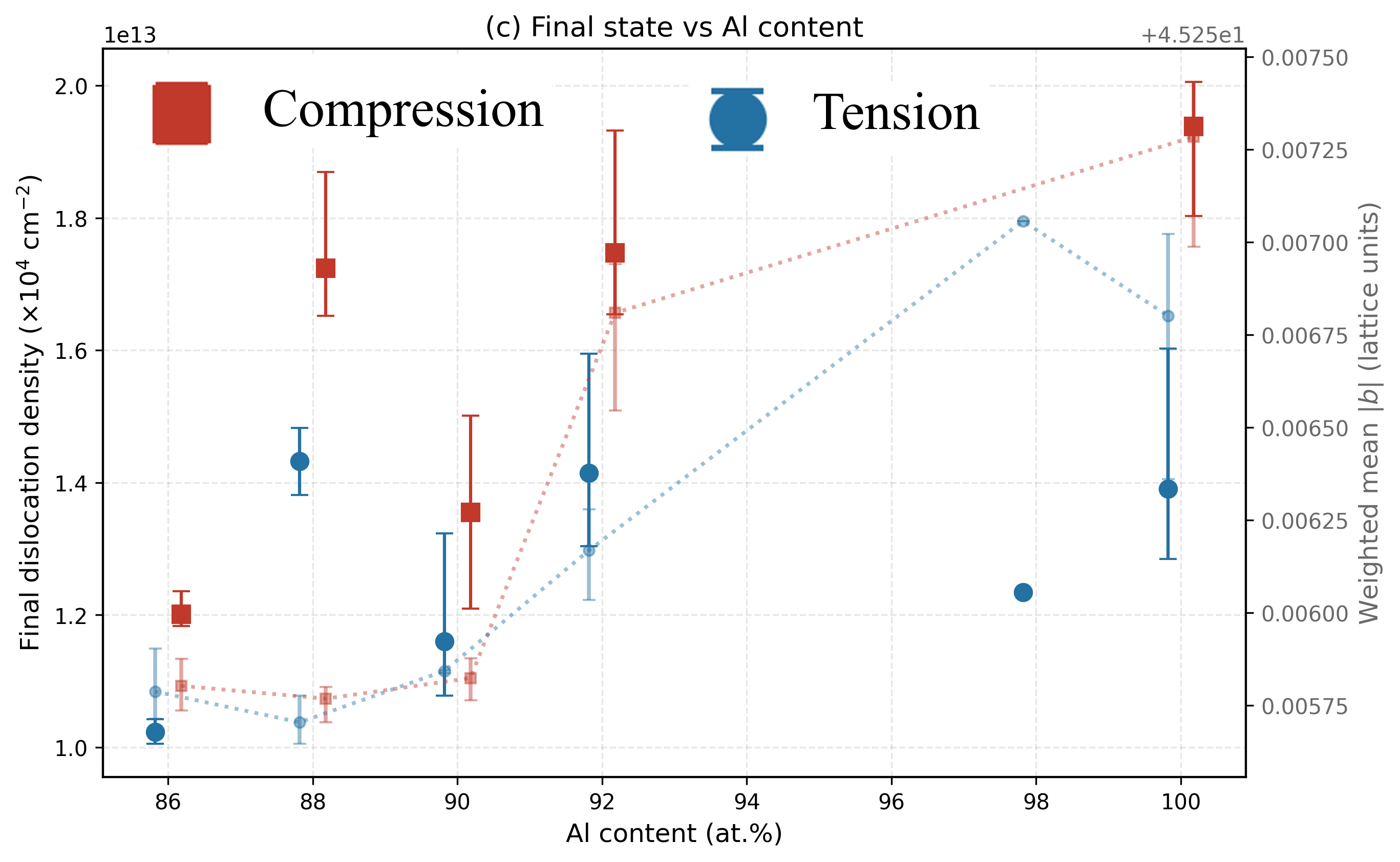} \caption{ Composition-dependent dislocation evolution in Al-rich alloys under tensile and compressive loading at a strain rate of $10^{8}$ s$^{-1}$. (a) Evolution of dislocation density as a function of absolute engineering strain. (b) Evolution of the weighted mean Burgers vector magnitude during deformation. (c) Final dislocation density and weighted mean Burgers vector magnitude as a function of Al content for tensile and compressive loading. Shaded regions and error bars represent variations among loading directions. } \label{fig:defect_evolution} \end{figure}

The evolution of dislocation structures during deformation is summarized in Figure~\ref{fig:defect_evolution}. All compositions exhibit an increase in dislocation density with increasing strain, indicating continuous defect nucleation and multiplication during plastic flow. Following an initial decrease at small strains, the dislocation density increases progressively as deformation proceeds, reaching values on the order of $10^{13}$ cm$^{-2}$ at the largest applied strains (Figure~\ref{fig:defect_evolution}(a)). The increase is most pronounced in the Al-rich compositions, particularly those approaching the pure-Al limit, which develop the largest final dislocation densities under both tensile and compressive loading. The observed trends indicate that alloy chemistry strongly influences the extent of defect generation during plastic deformation. Transition-metal-rich compositions maintain comparatively lower dislocation densities throughout deformation, whereas Al-rich alloys exhibit sustained dislocation multiplication and more extensive defect accumulation. These quantitative trends are consistent with the increasingly developed dislocation networks observed in the DXA snapshots presented in Figure~\ref{fig:ss}. In contrast to the strong composition dependence of dislocation density, the weighted mean Burgers vector magnitude remains nearly constant throughout deformation (Figure~\ref{fig:defect_evolution}(b)). Only minor fluctuations are observed with strain, indicating that the characteristic Burgers-vector magnitude changes little during loading. This behavior suggests that the mechanical response is governed primarily by changes in dislocation population and defect character rather than by systematic changes in Burgers-vector size. The final-state quantities are summarized in Figure~\ref{fig:defect_evolution}(c). Both tensile and compressive loading show a general increase in final dislocation density with increasing Al concentration, although the magnitude of the increase differs between loading modes. Compression \rev{produces} larger final dislocation densities than tension, indicating enhanced defect multiplication under compressive loading. \rev{To quantify this trend, the final dislocation densities were averaged over the independent solidification realizations and all three loading directions for each composition. Compression consistently produced higher mean dislocation densities than tension, although the magnitude of the difference varied with alloy composition. The associated standard deviations, shown as shaded bands in Figure~\ref{fig:defect_evolution}, demonstrate that the compression--tension difference exceeds the statistical scatter for most compositions. A paired statistical comparison between matched tension and compression simulations further confirmed that the average increase in dislocation density under compression is significant. The abstract and discussion have been revised accordingly to describe this trend as an average behaviour across the composition range rather than as a universal feature of every individual composition.} The corresponding Burgers-vector magnitudes show only weak composition dependence, further supporting the conclusion that alloy chemistry primarily influences the accumulation and distribution of deformation defects. Overall these results demonstrate that alloy composition strongly affects the evolution of defect populations during deformation. As the transition-metal content decreases and the alloy approaches the pure-Al limit, deformation is accompanied by substantially greater dislocation accumulation. When combined with the DXA observations in Figure~\ref{fig:ss}, these results indicate the development of increasingly extensive dislocation networks in the Al-rich compositions, providing a mechanistic basis for the observed composition dependence of mechanical behavior.

\newpage  \newpage

\subsection{Interface accommodation operators}
\label{sec:operators}

Sections~3.1 and~3.2 established the composition-dependent deformation behavior of the Al-rich alloys and showed that plastic flow is accompanied by substantial dislocation activity. Those results, however, do not explain how strain is accommodated within the intermetallic network itself. Experimental observations of deformed Al$_{92}$(TiFeCoNi)$_2$ reveal dislocation activity within Al$_3$Ti, stacking faults within Al$_9$(Fe,Co,Ni)$_2$, and strongly curved intermetallic lamellae, indicating that heterophase interfaces are active participants in deformation rather than passive boundaries \cite{Shang2024NatComm}. To quantify this interfacial response, atomistically resolved models of Al/Al$_3$Ti, Al/Al$_9$M$_2$, and Al$_9$M$_2$/Al$_3$Ti were subjected to tensile, compressive, and shear loading, and their mechanical and structural responses were analyzed using the accommodation framework introduced in Section~2. \rev{The bulk polycrystalline and bicrystal simulations address complementary aspects of the deformation behaviour and should therefore be interpreted together rather than compared directly. The bulk simulations identify the phase constitution that develops during directional solidification, establish the dominant bulk deformation mechanisms, and determine the phase pairs most frequently present in the intermetallic network. Consequently, the Al/Al$_3$Ti, Al/Al$_9$M$_2$, and Al$_9$M$_2$/Al$_3$Ti interfaces examined here are selected directly from the simulated and experimentally observed microstructures rather than from idealized crystallographic combinations. The bulk simulations further demonstrate that plastic deformation is dominated by Shockley partial dislocations in the Al-rich matrix and that dislocation accumulation decreases with increasing transition-metal content. These observations provide the bulk reference state against which the contribution of individual heterophase interfaces can be evaluated.} \rev{The bicrystal simulations isolate the response of each interface under well-defined loading conditions. In the bulk microstructure, the response of an individual interface is inseparable from the effects of neighboring interfaces, grain boundaries, triple junctions, and heterogeneous stress distributions within the surrounding matrix. Constructing isolated bicrystals removes these complexities and permits direct measurement of interface broadening, roughening, and migration. The resulting accommodation operators therefore characterize the intrinsic structural response of each interface type under controlled conditions. They are not intended to predict the total accommodation of the interconnected intermetallic network, but instead provide interface-specific quantities that can be incorporated into future mesoscale descriptions of the heterogeneous microstructure.} Figure~\ref{fig:interface_strength} summarizes the mechanical response of the three representative interfaces. Under compressive loading, shown in Figure~\ref{fig:interface_strength}(a), the interfaces exhibit pronounced differences in post-yield behavior. The Al$_9$M$_2$/Al$_3$Ti interface shows comparatively smooth flow after yielding, whereas Al/Al$_3$Ti and Al/Al$_9$M$_2$ display stronger stress fluctuations, consistent with intermittent structural rearrangements. Under tensile loading, Figure~\ref{fig:interface_strength}(b) shows that all three interfaces deform elastically before reaching peak stresses in the GPa range, but the Al$_9$M$_2$/Al$_3$Ti interface sustains the highest peak stress and then relaxes more gradually than the other two systems. The shear response in Figure~\ref{fig:interface_strength}(c) is even more anisotropic. Shear XY, XZ, and YZ activate different deformation pathways and produce distinct peak stresses and post-yield responses, demonstrating that interfacial deformation depends strongly on crystallographic orientation. The corresponding 0.2\% offset yield strengths are summarized in Figure~\ref{fig:interface_strength}(d). Yield strengths vary from very low values under tensile loading of Al$_9$M$_2$/Al$_3$Ti to substantially larger values under shear loading of Al/Al$_3$Ti, confirming that interfacial resistance is highly direction dependent and cannot be represented by a single scalar quantity. \rev{The 0.2\% offset criterion is adopted solely to provide a consistent basis for comparing the different interface and loading combinations. The corresponding peak stresses produce the same ordering of interface strength, indicating that the relative ranking is not sensitive to the particular yield definition employed.}

\begin{figure}[H]
  \centering
  \includegraphics[width=\linewidth]{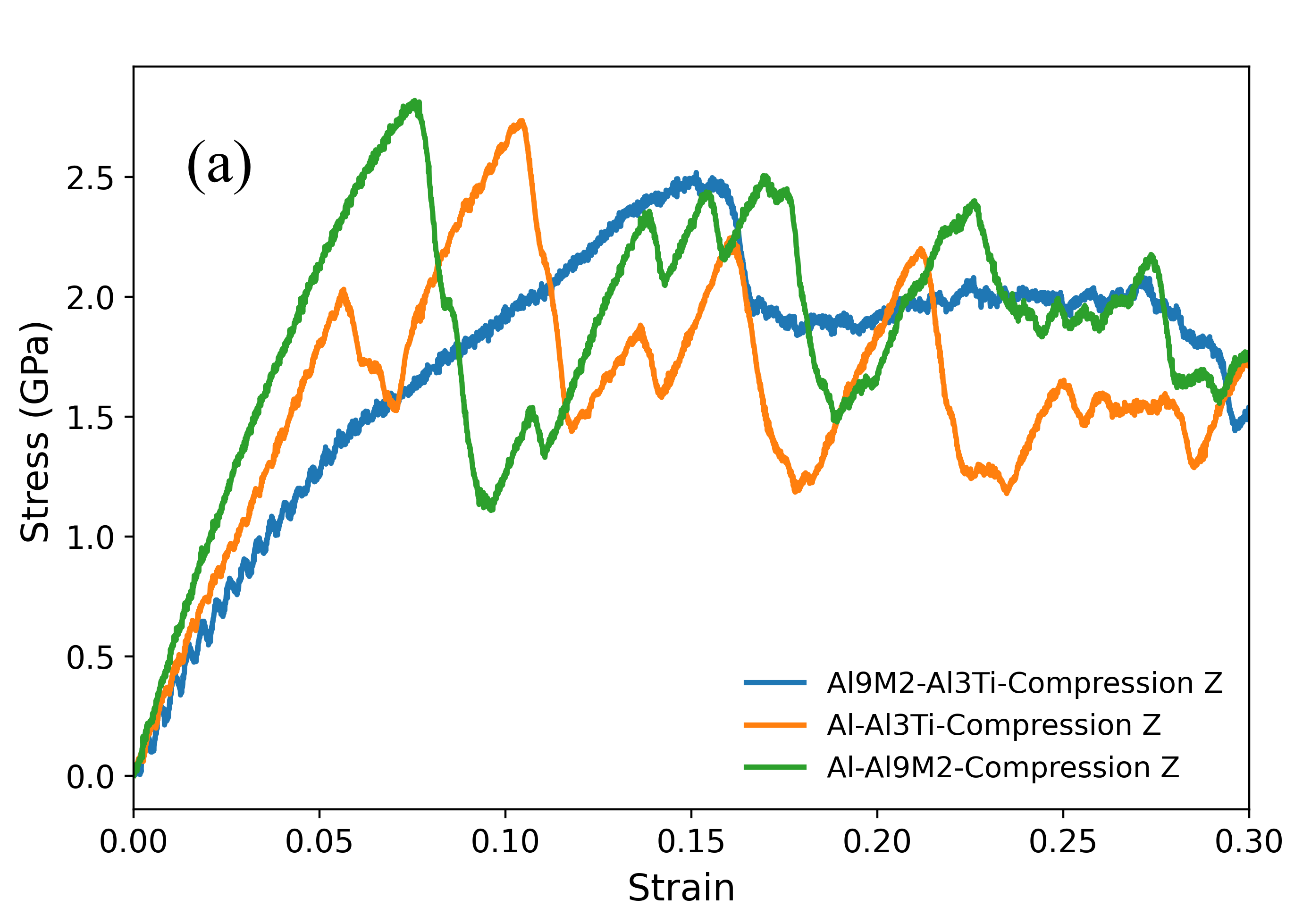}
  \includegraphics[width=\linewidth]{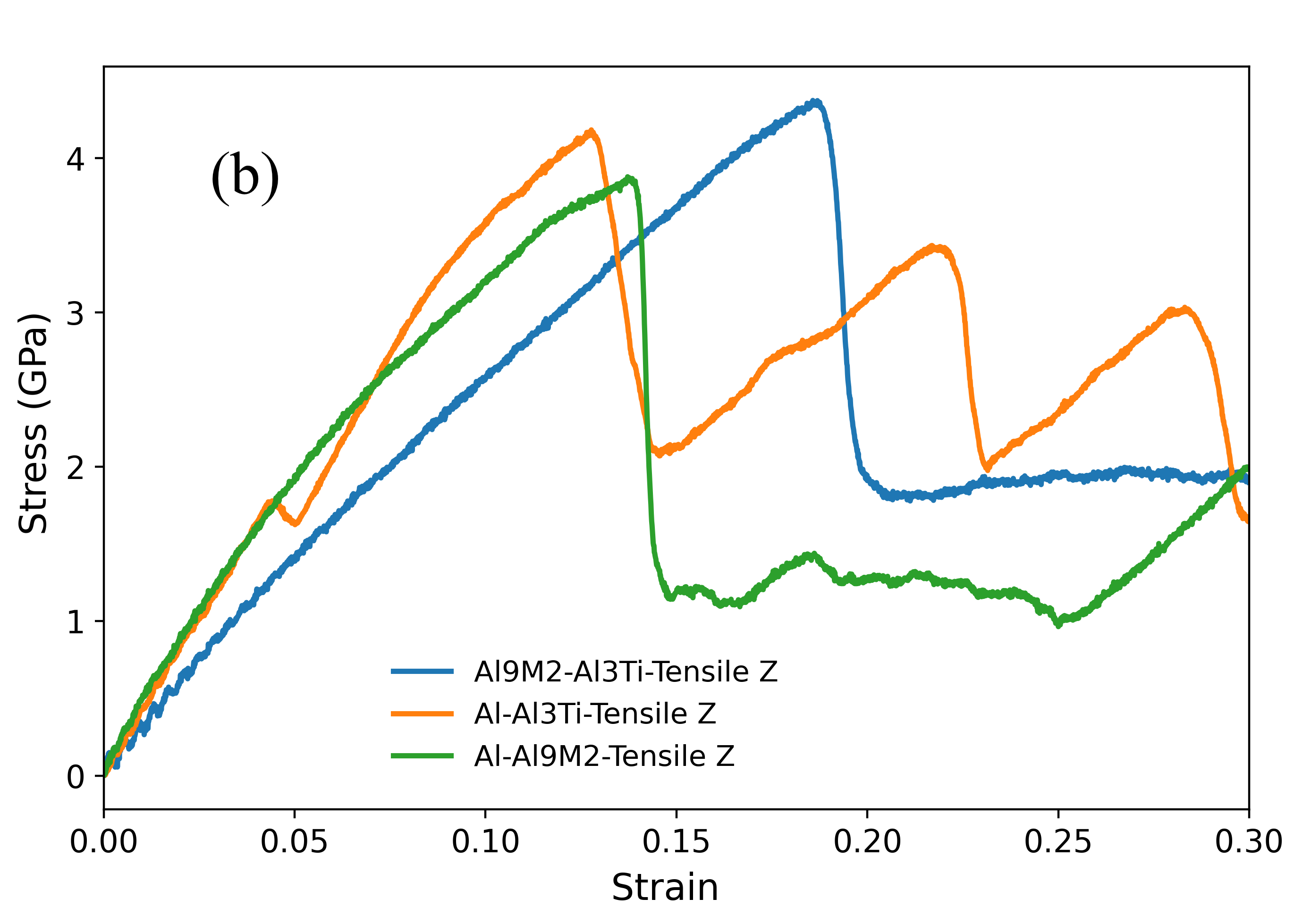}
\end{figure}

\begin{figure}[H]
  \centering
  \includegraphics[width=\linewidth]{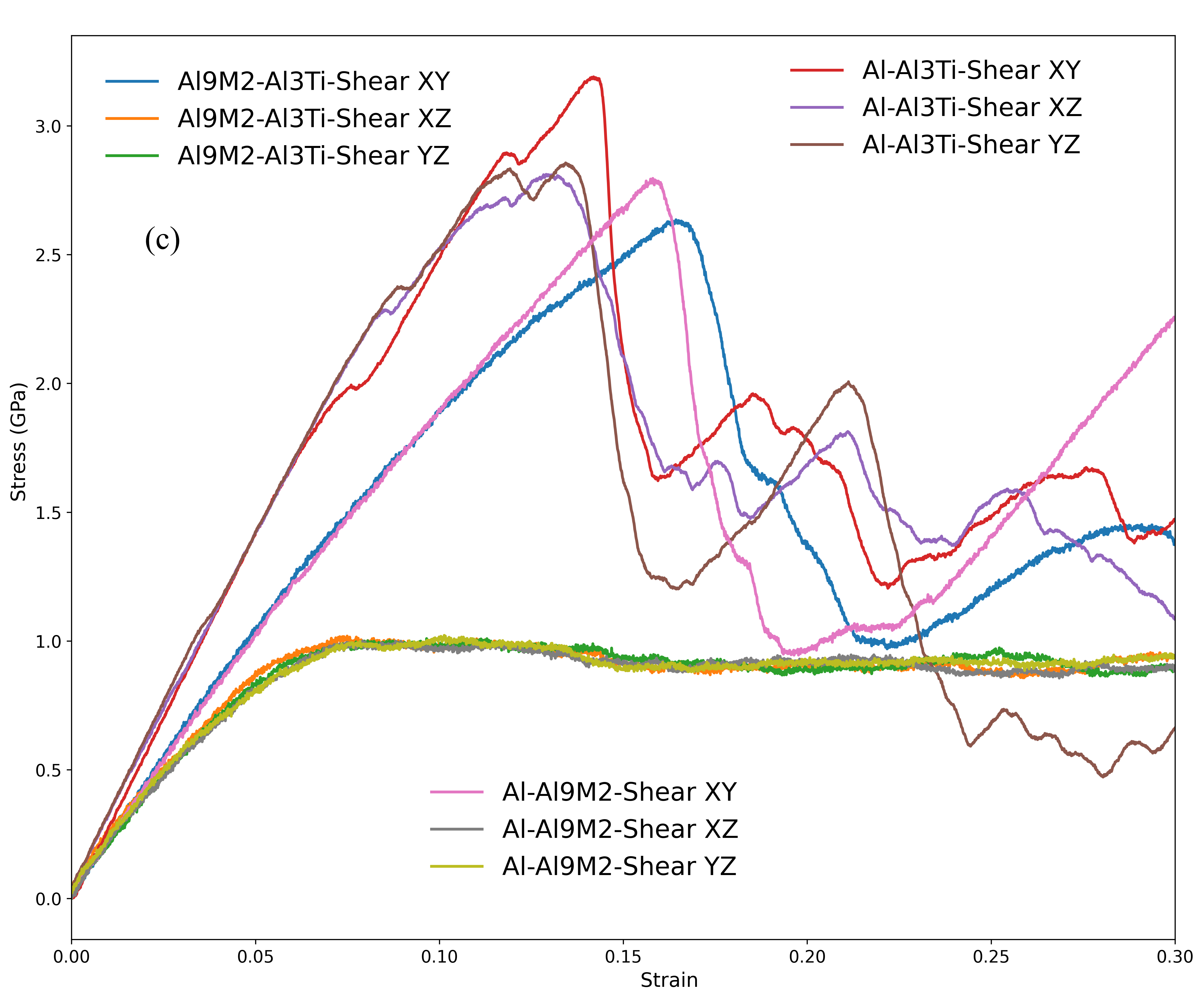}
  \includegraphics[width=\linewidth]{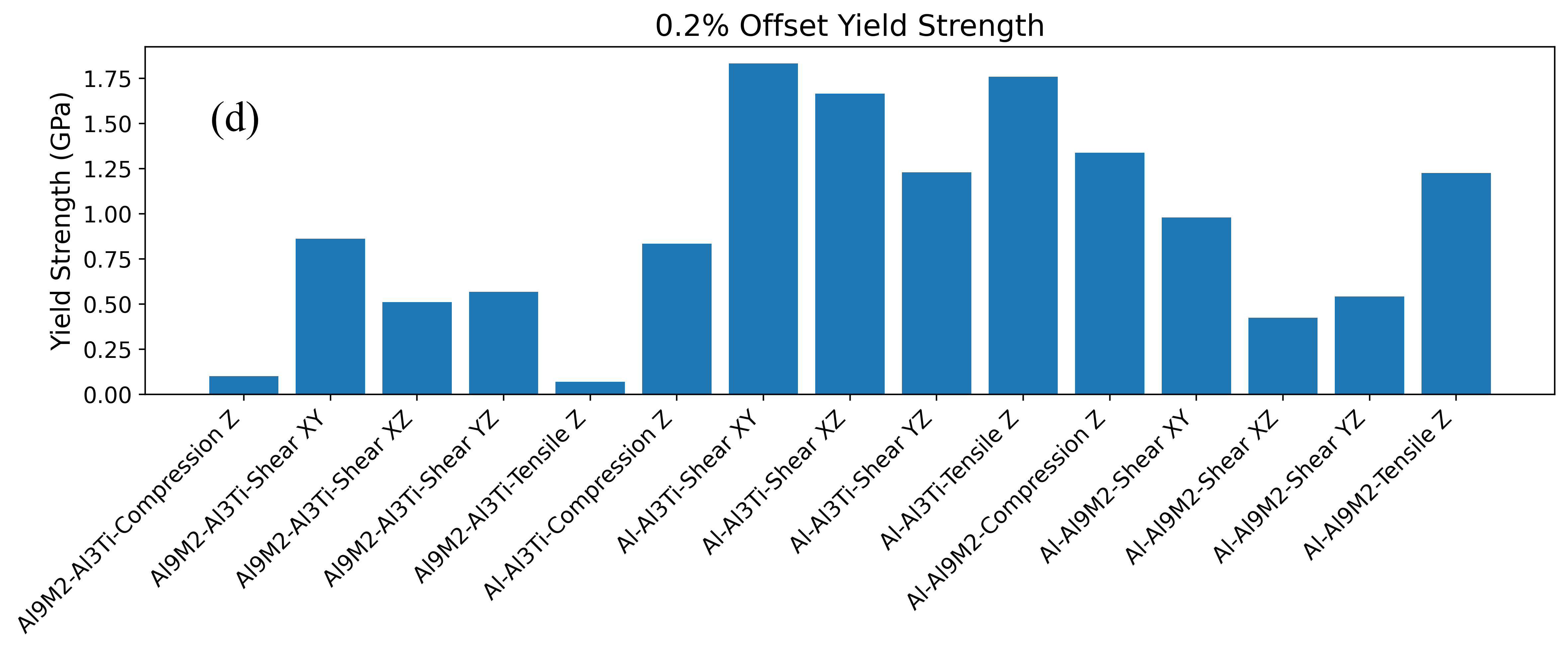}
  \caption{Mechanical response of heterophase interfaces under uniaxial and shear loading. (a) Compressive, and (b) tensile (z-direction) stress--strain curves for the Al$_9$M$_2$/Al$_3$Ti, Al/Al$_3$Ti, and Al/Al$_9$M$_2$ interface models. (c) Shear stress--strain curves. (d) 0.2\% offset yield strength for all interface--loading combinations, showing strong anisotropy in interfacial resistance.}
  \label{fig:interface_strength}
\end{figure}

Mechanical strength alone does not capture the full interfacial response. To quantify strain accommodation, the evolution of $W$, $R$, and $\Delta z$ were monitored during deformation. Figure~\ref{fig:interface_operators}(a) shows the strain dependence of $W$ and $R$ for all three interfaces under the five loading modes. At small cumulative strain, all interfaces remain comparatively stable and exhibit only limited structural evolution. Beyond this early regime, the response diverges strongly by interface type and loading mode. The Al$_9$M$_2$/Al$_3$Ti interface shows the largest broadening and roughening under tensile loading, whereas the Al/Al$_3$Ti interface displays moderate evolution under compression and selected shear modes. By contrast, the Al/Al$_9$M$_2$ interface remains comparatively narrow and smooth over much of the loading history. These trends indicate that the structural response of an interface is not determined by peak strength alone: two interfaces may sustain similar stresses while evolving very differently at the atomic scale.

The cumulative structural response is condensed into the Interface Accommodation Index,
\begin{equation}
A_{\mathrm{int}} = \Delta W + \Delta R + |\Delta z|,
\end{equation}
which combines interface broadening, roughening, and migration into a single measure of accommodation capacity. The resulting values are shown in Figure~\ref{fig:interface_operators}(b), with each bar decomposed into its three contributions. Several trends are clear. First, tensile loading consistently produces the largest accommodation response across all interfaces. Second, the Al$_9$M$_2$/Al$_3$Ti interface exhibits the highest overall accommodation capacity, reaching $A_{\mathrm{int}} \approx 17.4$~\AA\ under tension. Third, broadening is the dominant contribution in the highest-accommodation cases, while roughening and migration provide smaller but still physically meaningful contributions. The migration trajectories in Figure~\ref{fig:interface_operators}(c) reinforce this picture: Al$_9$M$_2$/Al$_3$Ti under tension shows monotonic migration, whereas migration under shear remains much more limited and often saturates quickly. The Al/Al$_3$Ti interface exhibits larger migration amplitudes than some other cases, but its total accommodation remains lower because broadening and roughening are less pronounced. The Al/Al$_9$M$_2$ interface shows the smallest overall structural evolution, indicating a more limited capacity to redistribute strain through interface restructuring.

\begin{figure}[H]
  \centering
  \includegraphics[width=\linewidth]{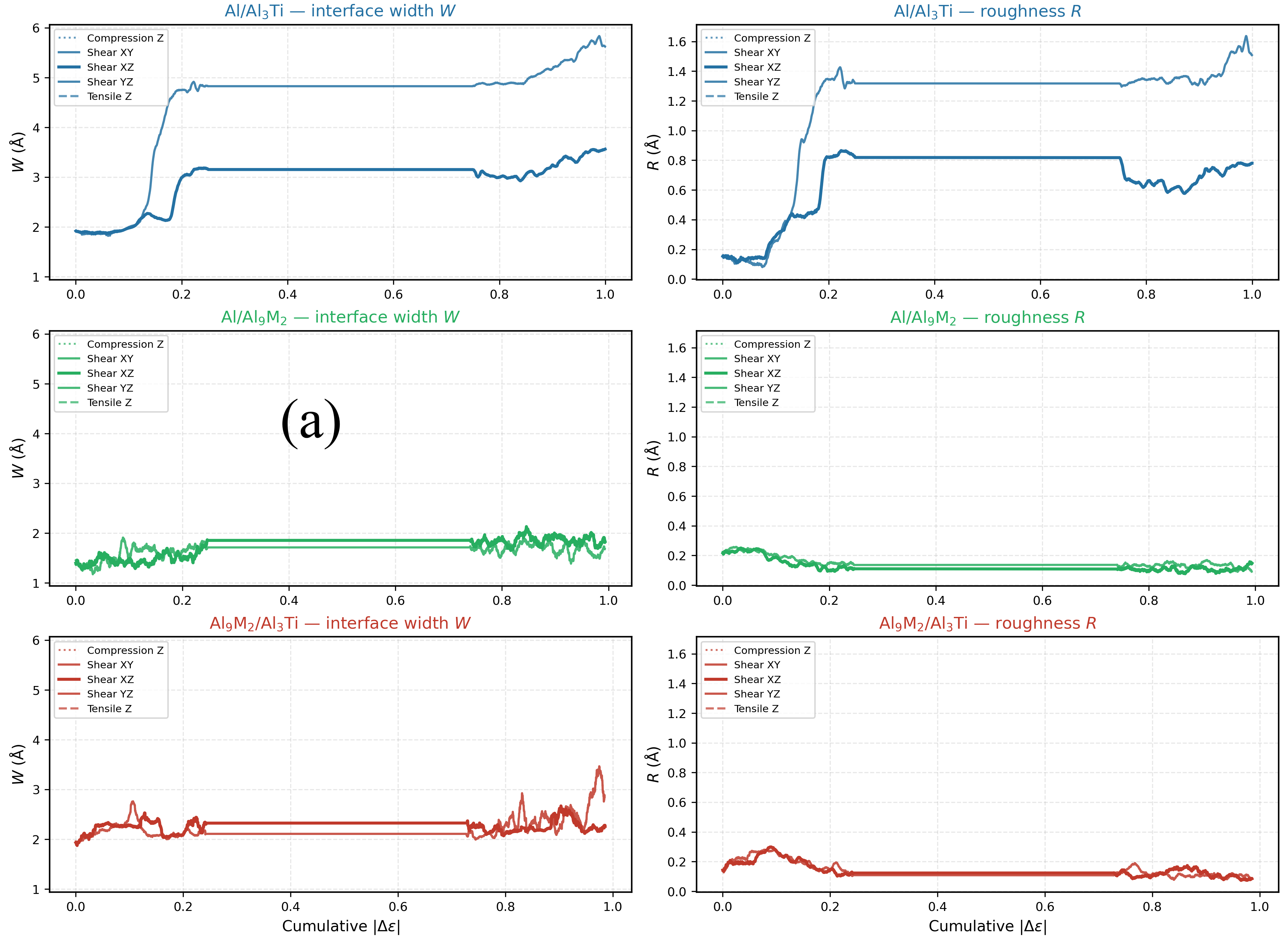}
  \includegraphics[width=\linewidth]{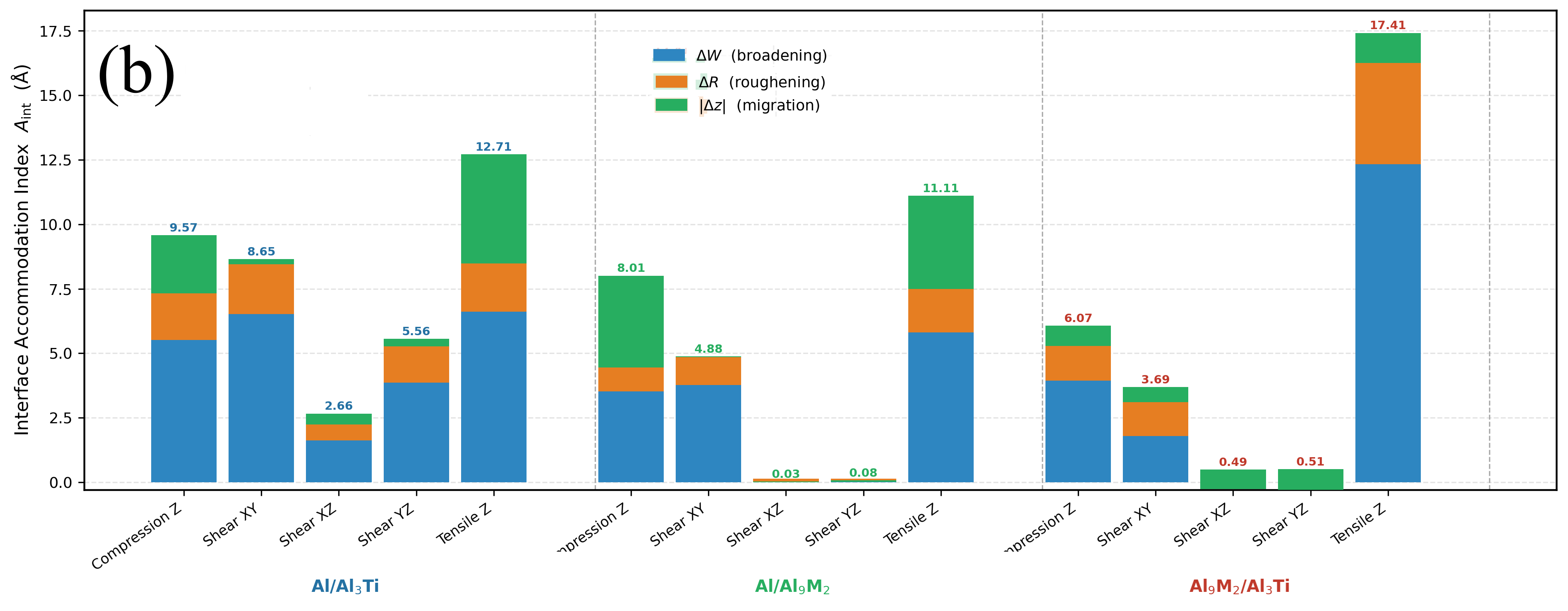}
\end{figure}

\begin{figure}[H]
  \centering
  \includegraphics[width=\linewidth]{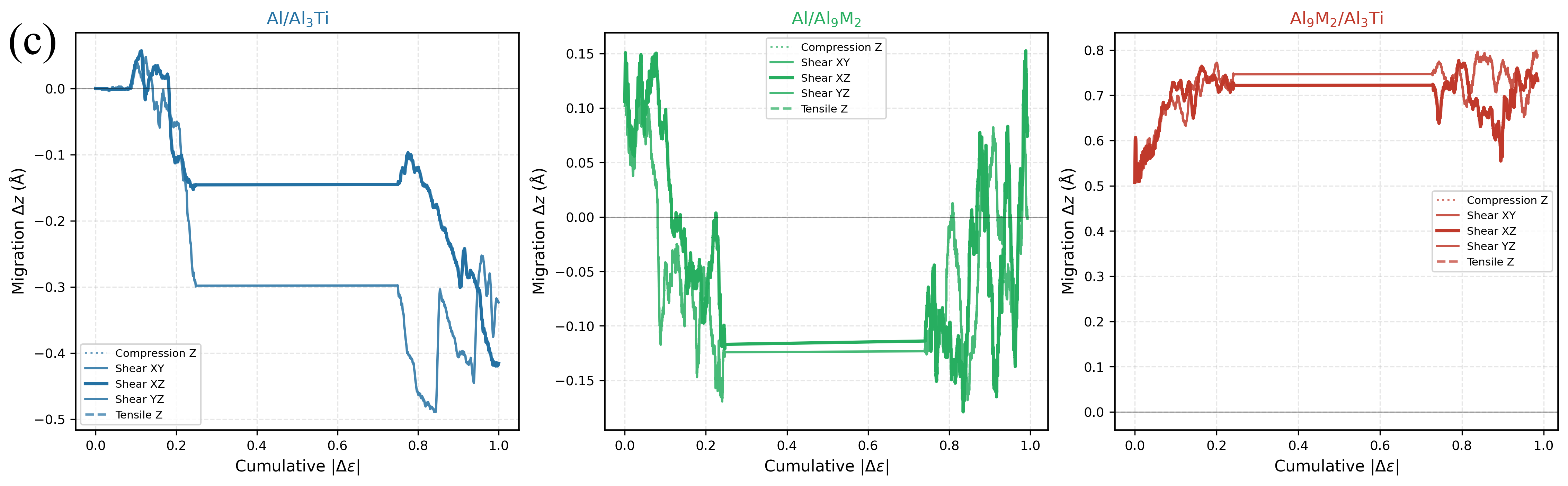}
  \caption{Interface accommodation operators. (a) Evolution of interface width $W(\varepsilon)$ and roughness $R(\varepsilon)$ for the three interface systems under all loading modes. (b) Interface Accommodation Index $A_{\mathrm{int}} = \Delta W + \Delta R + |\Delta z|$ for each interface and loading direction, decomposed into broadening, roughening, and migration contributions. (c) Net interface migration $\Delta z(\varepsilon)$ as a function of cumulative strain.}
  \label{fig:interface_operators}
\end{figure}

The accommodation results suggest that interface strength alone is not sufficient to describe deformation behavior. The Al$_9$M$_2$/Al$_3$Ti interface exhibits the largest accommodation response, particularly under tensile loading, despite remaining one of the stronger interfaces in the study. This indicates that a strong interface can also be highly accommodating if it is capable of evolving through broadening, roughening, and migration during deformation. The structural changes observed in Figure~\ref{fig:interface_operators} show that heterophase interfaces are active participants in strain accommodation. Rather than acting solely as barriers to dislocation motion, they can absorb deformation through local restructuring and redistribute strain over a larger volume. Interfaces that exhibit larger accommodation responses are therefore likely to play an important role in delaying damage accumulation and improving ductility in intermetallic-rich alloys. The connection between interface accommodation and ductility is explored in the next section.

\subsection{Interface accommodation pathways and implications for ductility}
\label{sec:pathways}

The accommodation metrics extracted from the molecular dynamics simulations provide a quantitative way to compare how effectively different interfaces redistribute strain during deformation. Table~\ref{tab:operators} summarizes representative accommodation responses for the interface--loading combinations examined in this work.

\begin{table}[H]
\centering
\caption{Accommodation metrics for representative interface--loading conditions. The table reports the contributions of interface broadening ($\Delta W$), roughening ($\Delta R$), and migration ($|\Delta z|$) to the accommodation index, together with the initial interface width $W_0$, initial roughness $R_0$, and the normalized accommodation parameter $D_{\mathrm{int}}=\Delta W/W_0+\Delta R/R_0$. All lengths are reported in \AA; $D_{\mathrm{int}}$ is dimensionless.}
\label{tab:operators}

\footnotesize
\setlength{\tabcolsep}{4pt}
\renewcommand{\arraystretch}{1.15}

\begin{tabular}{llccccccc}
\hline
Interface & Loading & $\Delta W$ & $\Delta R$ & $|\Delta z|$ & $A_{\mathrm{int}}$ & $W_0$ & $R_0$ & $D_{\mathrm{int}}$\\
\hline

Al$_9$M$_2$/Al$_3$Ti & Tension     & 12.3 & 3.9 & 1.2 & 17.4 & 96.80 & 1.18 &
\multirow{7}{*}{\centering Eq.~(\ref{eq:Dint})} \\

Al/Al$_3$Ti          & Tension     & 6.6 & 1.9 & 4.2 & 12.7 & 4.63 & 0.38 & \\

Al/Al$_3$Ti          & Compression & 5.5 & 1.8 & 2.3 & 9.6  & 4.63 & 0.38 & \\

Al/Al$_9$M$_2$       & Compression & 3.5 & 0.9 & 3.6 & 8.0  & 5.29 & 0.47 & \\

Al$_9$M$_2$/Al$_3$Ti & Shear XY    & 1.8 & 1.3 & 0.6 & 3.7  & 96.80 & 1.18 & \\

Al$_9$M$_2$/Al$_3$Ti & Shear XZ    & 0.2 & 0.1 & 0.7 & 1.0  & 96.80 & 1.18 & \\

Al$_9$M$_2$/Al$_3$Ti & Shear YZ    & 0.3 & 0.0 & 0.8 & 1.1  & 96.80 & 1.18 & \\

\hline
\end{tabular}
\end{table}

\rev{The addition of the migration column provides a more complete decomposition of the accommodation index by showing the relative contribution of each accommodation operator. Although Al$_9$M$_2$/Al$_3$Ti under tension and Al/Al$_3$Ti exhibit comparable values of $A_{\mathrm{int}}$, their accommodation mechanisms differ markedly. Accommodation at the Al$_9$M$_2$/Al$_3$Ti interface is dominated by interface broadening, with migration contributing comparatively little, whereas the Al/Al$_3$Ti interface exhibits a substantially larger contribution from interface migration. This distinction highlights that similar overall accommodation capacities can arise through different combinations of structural accommodation mechanisms. The normalized parameter $D_{\mathrm{int}}$ was introduced to facilitate comparison between interfaces with different initial morphologies. By normalizing the broadening and roughening contributions by their respective initial values, $D_{\mathrm{int}}$ reduces the influence of the initial interface width and roughness on the measured accommodation response. Comparison of the $A_{\mathrm{int}}$ and $D_{\mathrm{int}}$ rankings therefore demonstrates that the relative ordering of interface accommodation is not an artifact of the reference configuration, but remains robust after accounting for differences in the initial interface structure.}

\rev{Before drawing conclusions from these numbers, the robustness of the ranking to the definition of the index was tested using the weighted family of Eq.~(\ref{eq:IAI_weighted}), as described in Section~\ref{sec:sensitivity}. Table~\ref{tab:sensitivity} reports, for each weighting variant, the condition ranked most accommodating and the Spearman rank correlation $\rho$ with the equal-weight ordering. The ranking is essentially unchanged under all weightings in which broadening or roughening carries appreciable weight ($\rho \geq 0.93$), and the separation between tensile and shear loading is preserved in every variant without exception. The two migration-weighted variants reorder the top of the list, promoting Al/Al$_3$Ti above Al$_9$M$_2$/Al$_3$Ti, with the migration-only limit departing furthest from the baseline ($\rho = 0.64$). As set out in Section~\ref{sec:sensitivity}, this reflects a genuine difference in accommodation pathway between the two boundaries rather than an instability of the metric: Al/Al$_3$Ti translates while remaining comparatively sharp, whereas Al$_9$M$_2$/Al$_3$Ti disorders in place. It is precisely this distinction that motivates the use of a composite accommodation index rather than any single operator. Accordingly, the principal finding of this work, namely that the Al$_9$M$_2$/Al$_3$Ti interface under tensile loading exhibits the greatest structural accommodation, remains robust provided the index retains sensitivity to interface broadening and roughening. This behavior is captured only by the composite accommodation measure and is not reproduced when interface migration is considered in isolation.}

\rev{
\begin{table}[H]
\centering
\caption{\rev{Sensitivity of the interface ranking to the definition of the accommodation index. $\rho$ is the Spearman rank correlation between the ordering produced by each variant and that produced by the equal-weight index.}}
\label{tab:sensitivity}
\small
\begin{tabular}{lccc}
\hline
Index variant & $(w_W, w_R, w_z)$ & Highest-ranked condition & $\rho$ \\
\hline
Equal weight (baseline)   & $(1, 1, 1)$       & Al$_9$M$_2$/Al$_3$Ti, tension & 1.00 \\
Broadening-dominant       & $(2, 0.5, 0.5)$   & Al$_9$M$_2$/Al$_3$Ti, tension & 1.00 \\
Roughening-dominant       & $(0.5, 2, 0.5)$   & Al$_9$M$_2$/Al$_3$Ti, tension & 0.96 \\
Migration-dominant        & $(0.5, 0.5, 2)$   & Al/Al$_3$Ti, tension          & 0.93 \\
Broadening only           & $(3, 0, 0)$       & Al$_9$M$_2$/Al$_3$Ti, tension & 1.00 \\
Roughening only           & $(0, 3, 0)$       & Al$_9$M$_2$/Al$_3$Ti, tension & 0.93 \\
Migration only            & $(0, 0, 3)$       & Al/Al$_3$Ti, tension          & 0.64 \\
Normalized $D_{\mathrm{int}}$ & $\Delta W/W_0,\ \Delta R/R_0,\ \Delta z$ & Table 1 & Table 1 \\
\hline
\end{tabular}
\end{table}
}

Several observations emerge from the accommodation analysis. Tensile loading produces the largest accommodation response for all three interfaces. Among them, the Al$_9$M$_2$/Al$_3$Ti interface exhibits the highest accommodation capacity, reaching an accommodation index of approximately 17~\AA. Figure~\ref{fig:interface_operators}(b) further shows that this response is dominated by interface broadening, while roughening provides a secondary contribution and interface migration remains comparatively modest. These atomistic observations are consistent with the deformation features reported experimentally for Al$_{92}$(TiFeCoNi)$_2$, including curved intermetallic lamellae, stacking faults, and localized lattice distortions within the intermetallic colonies \cite{Shang2024NatComm}. Rather than behaving solely as barriers to dislocation motion, the heterophase interfaces evolve through broadening, roughening, and limited migration, thereby redistributing deformation over an extended interfacial region. Within the framework of disconnection theory \cite{HirthPond1996,Cahn2006,Han2018}, the measured structural evolution is consistent with interface-mediated defect activity, in which broadening reflects the development of a diffuse interfacial region, roughening reflects the accumulation of local steps and ledges, and migration corresponds to the net displacement of the interface. \rev{The accommodation framework developed here is formulated entirely in terms of measurable structural observables and is therefore not restricted to the present alloy system. Although demonstrated for a complex additively manufactured Al alloy containing multiple heterophase interfaces, the same methodology can be extended to other multiphase materials provided reliable atomistic descriptions of the constituent phases and interfaces are available. Because the accommodation operators are extracted directly from molecular dynamics trajectories, their quantitative values remain material specific and depend on the underlying interatomic potential together with the crystallographic characteristics of the interfaces considered. Application of the framework to a different material system therefore requires the corresponding atomistic description to be validated against first-principles calculations and available experimental data. Within these framework, the present methodology provides a transferable framework for quantifying and comparing interface accommodation across chemically and crystallographically distinct multiphase alloys.}

An important outcome of the present study is that accommodation capacity does not decrease with increasing interfacial strength. The Al$_9$M$_2$/Al$_3$Ti interface exhibits both relatively high mechanical resistance and the largest accommodation response under tensile loading. This combination suggests that intermetallic--intermetallic interfaces can simultaneously sustain load while accommodating local strain through structural evolution, providing deformation pathways that are not available in simpler matrix--intermetallic systems.

The results therefore \rev{are consistent with} an interface-mediated contribution to ductility in intermetallic-rich alloys\rev{, while stopping short of establishing one}. \rev{The proposed sequence in which interface broadening, roughening, and migration reduce local strain concentrations, delay void and crack initiation, and thereby extend the plastic deformation range is physically consistent with both the present simulations and the experimental observations. However, the latter stages of this sequence are not directly evaluated in the present work. Quantities such as fracture strain, work of separation, crack-tip shielding, and damage evolution were not calculated. Establishing these links will require simulations containing pre-existing flaws together with sufficiently large simulation cells to capture damage-process-zone development. Accordingly, the interface-accommodation mechanism proposed here should be regarded as a candidate explanation for the experimentally observed deformability of Al$_{92}$(TiFeCoNi)$_2$, while also providing a framework that can be tested and extended to other complex multiphase alloy systems.}

\rev{\subsection{Limitations and scope}}
\label{sec:limitations}

\rev{The present work establishes an atomistic framework for quantifying structural accommodation at heterophase interfaces under controlled loading conditions. The accommodation operators introduced here are defined directly from measurable structural observables and therefore provide a transferable methodology for comparing interface response across chemically and crystallographically distinct multiphase materials. Their quantitative values, however, remain specific to the material system under investigation and depend on the underlying atomistic description of the constituent phases and interfaces. Application of the framework to other alloy systems therefore requires the corresponding interatomic potentials and crystallographic models to be validated against first-principles calculations and available experimental observations. The bicrystal simulations isolate representative interface configurations to determine their intrinsic accommodation behavior independently of neighboring interfaces, grain boundaries, and heterogeneous stress fields. Consequently, the reported accommodation capacities should be interpreted as interface-specific quantities rather than direct predictions of the collective response of an interconnected intermetallic network. Likewise, only one representative orientation relationship was considered for each phase pair, and the quantitative values reported here should not be interpreted as universal for all interfaces of the same chemistry. These considerations define the scope of the present framework while providing a foundation for future multiscale investigations of interface-mediated deformation in complex multiphase alloys.}

%% ---------------------------------------------------------------
%%  4. CONCLUSIONS
%% ---------------------------------------------------------------
\section{Conclusions}

This work investigated deformation accommodation in intermetallic-rich Al alloys through large-scale molecular dynamics simulations of bulk microstructures and \rev{experimentally motivated} heterophase interfaces. Bulk deformation analysis showed that Shockley partial dislocations dominate plastic flow across all compositions, while alloy chemistry governs the evolution of dislocation networks and defect accumulation. Beyond these conventional bulk mechanisms, the simulations demonstrate that heterophase interfaces actively participate in deformation through interface broadening, roughening, and migration, with the accommodation response depending strongly on both interface chemistry and loading mode. To quantify these structural changes, an accommodation framework based on directly measurable interface descriptors was developed, providing a quantitative basis for comparing deformation responses across matrix--intermetallic and intermetallic--intermetallic interfaces. Among the interfaces examined, Al$_9$M$_2$/Al$_3$Ti exhibits the largest accommodation capacity, particularly under tensile loading, while simultaneously maintaining high resistance to yielding. \rev{Within the representative interfaces investigated, this demonstrates} that accommodation capacity and interfacial strength are not necessarily competing attributes. \rev{A weighting sensitivity analysis confirms that this conclusion is robust to the definition of the accommodation index, while the normalized parameter $D_{\mathrm{int}}$ demonstrates that the ranking is not an artefact of differences in the initial interface structure.} Interpreted within a \rev{disconnection-consistent} framework, interface broadening, roughening, and migration represent observable \rev{structural changes consistent with interface-mediated defect activity}; \rev{however, disconnection cores, Burgers vectors, and interfacial defect fluxes were not resolved explicitly, and the proposed interpretation should therefore be regarded as physically consistent with, rather than a direct demonstration of, disconnection-mediated plasticity.} \rev{The accommodation framework introduced in this work provides a transferable methodology for quantifying interface-mediated deformation in chemically and crystallographically complex multiphase alloys. Although the present study demonstrates the framework using an additively manufactured Al$_{92}$(TiFeCoNi)$_2$ alloy, its application to other material systems requires the corresponding atomistic descriptions, crystallographic interface models, and interatomic potentials to be validated for the constituent phases and interfaces. The simulations quantify structural accommodation at interfaces rather than macroscopic ductility, fracture, or damage evolution, and the proposed connection between interface accommodation and improved ductility should therefore be regarded as a hypothesis requiring further experimental and computational validation. Within this scope, the present results identify interface chemistry and crystallography as promising design variables for controlling deformation accommodation and provide a practical framework for connecting atomistic simulations with future mesoscale descriptions of interface-mediated deformation in complex multiphase alloys.}

%% ---------------------------------------------------------------
%%  CRediT STATEMENT
%% ---------------------------------------------------------------
\section*{CRediT authorship contribution statement}
\textbf{Avik Mahata}: Writing -- review \& editing, Writing -- original draft,
Visualization, Validation, Supervision, Software, Resources, Project
administration, Methodology, Investigation, Funding acquisition, Formal
analysis, Data curation, Conceptualization.

%% ---------------------------------------------------------------
%%  DECLARATION OF COMPETING INTEREST
%% ---------------------------------------------------------------
\section*{Declaration of competing interest}
The author declares that there are no known competing financial interests or
personal relationships that could have appeared to influence the work reported
in this paper.

%% ---------------------------------------------------------------
%%  ACKNOWLEDGEMENTS
%% ---------------------------------------------------------------
\section*{Acknowledgements}
This work was supported by the National Science Foundation through ACCESS
supercomputing allocations MAT250103 and MAT240094.  Additional computational
resources were provided by Argonne National Laboratory under the Director's
Discretionary allocation for the project GNNMD.  Computational resources were
also provided by a National Science Foundation MRI Award granted to Wilkes
University (Award 1920129).

%% ---------------------------------------------------------------
%%  REFERENCES
%% ---------------------------------------------------------------
\bibliographystyle{elsarticle-num}
\bibliography{references}

\end{document}